\documentstyle[aps,bezier,epsfig,amsmath,prb,graphicx,amssymb]{revtex}

\renewcommand{\epsilon}{\varepsilon}
\renewcommand{\phi}{\varphi}

\newcommand{\be}{\begin{equation}}
\newcommand{\ee}{\end{equation}}

\begin{document}

\draft
\twocolumn[\hsize\textwidth\columnwidth\hsize\csname @twocolumnfalse\endcsname
\title{Geometrical Aspects of Aging and Rejuvenation in the Ising Spin Glass: 
A Numerical Study}

\author{Ludovic Berthier$^1$ and Jean-Philippe Bouchaud$^2$}

\address{$^1$Theoretical Physics, 1 Keble Road, Oxford, OX1 3NP, UK}

\address{$^2$Service de Physique de l'\'Etat Condens\'e,
Orme des Merisiers, 91191 Gif-sur-Yvette Cedex, France}

\date{\today}

\maketitle

\begin{abstract}
We present a comprehensive study of non-equilibrium phenomena in the
low temperature phase of the Edwards-Anderson Gaussian spin glass in 3 and 4 
spatial dimensions. Many effects can be understood in terms of a time
dependent coherence length, $\ell_T$, such that length scales smaller that 
$\ell_T$ are equilibrated, whereas larger length scales are essentially frozen.
The time and temperature dependence of $\ell_T$ is found to be compatible with
critical power-law dynamical scaling for small times/high temperatures, 
crossing over to an activated 
logarithmic growth for longer times/lower temperatures, in agreement with 
recent experimental results.
The activated regime is governed by a `barrier exponent' $\psi$ which we
estimate
to be $\psi \sim 1.0$ and $\psi \sim 2.3$ in 3 and 4 dimensions, 
respectively. We observe for the first time the rejuvenation and 
memory effects in the four dimensional sample, which, we argue, is unrelated 
to `temperature chaos'. Our discussion in terms of 
length scales allows us to address several experimentally 
relevant issues, such as super-aging versus sub-aging effects, 
the role of a finite cooling rate, or the so-called Kovacs effect.
\end{abstract}  



\vskip2pc]

\narrowtext

{\it But it's better to have little free time, then memories don't intrude. 
Still, my God, what an amazing phenomenon these memories are.} \\
D. Shostakovich (Letter to I. Glikman).

\section{Introduction}

Spin glasses represent a model system for more `complex' glassy materials.
As such, it has attracted a large attention in the last decades~\cite{young}. 
Despite the large number of theoretical, numerical and
experimental papers published in the field, 
the original spin glass problem is still far from being 
quantitatively understood~\cite{JP}.

Experimental facts that have to be explained are mostly of 
dynamical nature. The low-temperature dynamics of various spin glass systems 
has been thoroughly investigated~\cite{review_manip1,review_manip2}, and
a number of characteristic features have emerged.
Among these are the well-known aging behaviour of the thermoremanent 
magnetisation or the a.c. susceptibility in isothermal 
experiments~\cite{aging},  
and the spectacular `memory' and `rejuvenation' effects 
observed in temperature-shifts protocols~\cite{shift1,shift2,shift3,cycle3}.  

It is fair to say 
that none of the existing theories provides
a complete quantitative description of spin glass dynamics~\cite{review_aging}.
In a recent paper~\cite{JPB}, 
following earlier ideas put forward 
in the context of the droplet model~\cite{BM,drop,hajime},
the notion of separation
of length and time scales was argued to be crucial 
to account for the whole of experimental data. An important ingredient
is the existence of a time and temperature
dependent {\it coherence length}, $\ell_T(t_w)$, 
such that length scales smaller than $\ell_T$ are {\it equilibrated}, 
whereas length scales larger than $\ell_T$ are {\it frozen}.
As a result~\cite{JPB,BM,drop,hajime,babe}, 
the aging dynamics after a certain age $t_w$ is 
ascribed to the motion of objects of size $\sim \ell_T(t_w)$. 
Note that we do not need to
specify the topological nature of these objects, postulated to be 
compact droplets in
Ref.~\cite{drop}, 
but that could be also fractal, sponge-like 
structures~\cite{OMartin,Palassini-Young,lamarcq}.
A simple realization of the above scenario was recently
worked out in the context of the 2D XY model~\cite{BerthierEPL}, 
suggesting that systems with quasi-long range order~\cite{feldman} 
have a dynamics very similar to what is observed in spin glass experiments.

\begin{figure}[b]
\begin{center}
\psfig{file=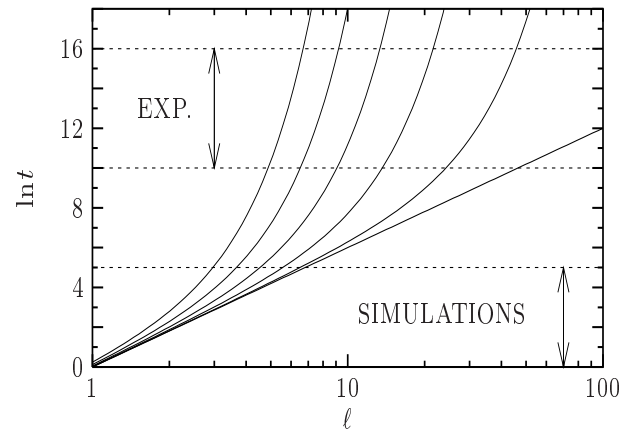,width=8.6cm,height=6.5cm} 
\end{center}
\caption{Growth
law of the coherence length, Eq.~(\ref{ellt}),
for $T/T_c = 1.0$, 0.9, $\cdots$ 0.5 (from bottom to top).
The numerical values are extracted from experiments \protect\cite{JPB}.
The separation of length scales within a given time 
window is very clear from these curves. 
Note also the crossover from the critical (power law) regime to the activated
regime, where the coherence length is stuck to a value of a few $\xi(T)$.} 
\label{growth}
\end{figure}

In this paper, following some previous 
works~\cite{hajime,huse,heiko,Marinari,review_simu}, 
we identify a coherence length that grows with the age of the system
in the Edwards-Anderson spin glass model~\cite{EA}.
We numerically test the growth law of $\ell_T$ proposed 
in Ref.~\cite{JPB}, according to which: 
\begin{equation}
t_w(\ell_T)  \simeq \tau_0 \ell_T^{z_c} \exp 
\left(\frac{\Upsilon(T) \ell_T^\psi}
{T} \right),
\label{ellt}
\end{equation}
where $z_c$ is the dynamical critical exponent, $\psi$ the 
so-called barrier exponent
that describes the growth of the energy barriers with length scales, 
and $\Upsilon(T)$
a temperature dependent free-energy scale that vanishes at the critical 
temperature $T_c$. This 
growth law, illustrated in Fig.~\ref{growth}, is motivated both by theoretical 
considerations~\cite{drop} and experimental results\cite{JPB}, and was 
used to analyze 
further experimental data\cite{Dupuis,Suedois-Yoshino}. 

By construction, this growth law reduces 
to usual critical scaling when the barriers on scale $\ell_T$ are 
much smaller than $T$. Assuming that the barrier scale behaves as
$\Upsilon(T) = \Upsilon_0 (1-T/T_c)^{\psi \nu}$ (where $\nu$ is the 
correlation length 
exponent~\cite{drop}, and $\Upsilon_0$ an energy scale of order $T_c$), the
crossover between critical scaling and activated scaling occurs for 
a dynamical crossover length $\xi(T)$ that diverges at $T_c$ as 
$(1-T/T_c)^{-\nu}$.
Remark that Eq.~(\ref{ellt}) is obviously not the only
possibility
to describe this crossover. However, this multiplicative form was found 
to represent
quite accurately the dynamics of the directed polymer or the Sinai 
model~\cite{Yoshinounpub,Marta}, where a
similar crossover between diffusive and activated dynamics takes 
place~\cite{PhysRep}.

As noted in Ref.~\cite{JPB}, the growth law Eq.~(\ref{ellt}) is difficult 
to distinguish, 
over a restricted range of length scales, 
from a pure power law $t_w \sim  \tau_0 \ell_T^{z}$ with a 
temperature dependent exponent $z=z(T) > z_c$. The latter was 
previously reported both 
numerically~\cite{hajime,heiko,Marinari,review_simu} and 
experimentally~\cite{zeeman}. 
However, we believe that Eq.~(\ref{ellt}) should be prefered.
One reason is that more elaborated experimental protocols,
such as temperature-shift experiments, reveal non-activated effects, 
as recalled below.
This non-activated behaviour is captured by Eq.~(\ref{ellt}), both
through the temperature dependent barrier term $\Upsilon(T)$ and the 
strong renormalisation of the
microscopic time scale by critical 
fluctuations~\cite{JPB,Dupuis,Suedois-Yoshino}. 

The present work is a quantitative investigation of 
the low-temperature, non-equilibrium dynamics of the Edwards-Anderson 
spin glass model~\cite{EA} in finite dimensions, $d=3$ and $d=4$.
It can be viewed as the numerical counterpart of Ref.~\cite{JPB}.
Here, we take advantage of the fact that simulations, 
unlike experiments~\cite{zeeman}, 
directly give access to $\ell_T$, to confirm some of the results obtained 
in Ref.~\cite{JPB} using indirect evidence.
To do so, we perform an extensive series of numerical experiments, 
including simple aging, temperature-shift and temperature-cycling 
protocols. We observe for the first time in this system 
the `rejuvenation and memory' and `Kovacs' effects, which 
are interpreted using the coherence length $\ell_T(t_w)$.
In turn, this allows us to shed new light on several questions
such as sub-aging effects, the issue of temperature chaos and the existence 
of an overlap length, and the very nature of the spin-glass phase. 
We emphasize also that although simulations and experiments are performed on
very different {\it time} windows, the {\it length} scales probed 
dynamically are actually not very different, see Fig.~\ref{growth}.

The paper is organized as follows.
Section~\ref{model} introduces the model and gives technical 
details on the simulation. 
Section~\ref{isothermal} focuses on simple (isothermal) aging.
The growth law of the coherence length is 
studied in section~\ref{growthsec}.
`Small' temperature-shift experiments are performed in section~\ref{Tshifts},
while `larger' shifts and cycles are studied in section~\ref{largesec}.
Physical implications of our results 
are discussed in section~\ref{physsec},
and section~\ref{discussion} summarizes and 
concludes the paper.

\section{Model and technical details}
\label{model}

We study the Edwards-Anderson spin glass 
model defined by the Hamiltonian~\cite{EA} 
\begin{equation}
H_J[\boldsymbol{s}] 
= - \sum_{\langle i j \rangle} J_{ij} s_i s_j,
\label{ham}
\end{equation}
where $\boldsymbol{s} = \{ s_i \}_{i=1,\cdots,N}$ are $N=L^d$ Ising spins 
located on a 3$d$ or 4$d$ (hyper)cubic lattice of linear size $L$, and
$J_{ij}$ are random variables taken from a Gaussian distribution
of mean 0 and variance 1. The sum is over nearest neighbors. 
The spin glass transition is believed~\cite{review_simu}
to take place at $T_c(d=3) = 0.95$ and $T_c(d=4)=1.8$.
In all this paper, the temperature is given in units of 
the critical temperature, $T/T_c(d) \to T$.

To study the aging dynamics, we use a rather large 
system linear size $L$, $30 \le L \le 40$ in $d=3$, 
and $15 \le L \le 26$ in $d=4$.
On the time scale of the simulation, the system never equilibrates
on a length scale larger than, say,  $\sim 8$ lattice spacings, and
we thus always work in the regime $\ell_T(t) \ll L$.
The dynamics associated to the Hamiltonian (\ref{ham}) is a
standard Monte Carlo algorithm, where the spins 
are randomly sequentially  updated. One Monte Carlo step represents $N$ 
attempts to 
update a spin. 

The behavior of the system is analyzed through the measurements 
of various physical quantities.
\begin{itemize}
\item We compute the energy density, defined by 
\begin{equation}
e = \frac{1}{N} \overline {\langle H_J[\boldsymbol{s}] \rangle},
\end{equation}
where $\langle \cdots \rangle$ stands for an average over
initial conditions and $\overline{\phantom{|}\cdots\phantom{|}}$
over the disorder.

\item We measure the two-time autocorrelation function defined by
\begin{equation}  
C(t_w+t,t_w) = \frac{1}{N} \sum_{i=1}^N  
\overline{\langle s_i(t_w+t) s_i(t_w) \rangle}.
\end{equation}
We will also consider an a.c. susceptibility like quantity 
$\chi(\omega,t_w)$, defined as~\cite{hajime}:
\begin{equation}\label{chidef}
\chi(\omega,t_w) \equiv  \frac{1- C(t_w+\frac{1}{\omega},t_w)}{T}.
\end{equation}
\item As in previous studies, we extract a coherence length by studying 
the dynamical 4-point correlation function, defined as
\begin{equation}
C_4 (r,t_w) = \frac{1}{N}
\sum_{i=1}^N  \overline{ \langle s_i^a(t_w) s_{i+r}^a(t_w) 
s_i^b (t_w) s_{i+r}^b(t_w) \rangle},
\label{c4r}
\end{equation}
where $(a,b)$ are two copies of the system starting 
from different initial conditions and evolving with independent 
thermal histories. 
This four-point correlation function
can be interpreted as the probability that
two spins separated by a distance $r$ have the same {\it relative} 
orientation  in two independent systems after time $t_w$, as 
is measured by a two-point function in a pure ferromagnet~\cite{bray}.

\end{itemize}

Our data are typically averaged over 15 (autocorrelation functions, 
energy density) to 50 (4-point correlation functions) 
realizations of the disorder.
Our data are thus reported without errorbars, which are typically
extremely small. 
Technically, the most difficult part of our work 
is then to perform a meaningful analysis of the data
in order to extract quantitative values of 
the various physical parameters.

\section{Isothermal aging: basic facts}
\label{isothermal}

In this section, we consider isothermal aging protocols. The system is quenched
at initial time $t_w=0$ from an {\it infinite} temperature
to a low temperature $T<1.0$ where it slowly evolves towards its 
equilibrium state. Although the phenomenology is very-well 
known~\cite{aging,review_aging},
and has already been thoroughly investigated in 
simulations~\cite{hajime,heiko,review_simu,Ricci96},
some important points are still poorly understood.
We discuss all these aspects in some details in this section. 
We evaluate, in particular, 
the implications of Eq.~(1) for the theoretical description of the data.

\subsection{The spin-spin correlation function: general considerations}

\begin{figure}
\begin{center}
\psfig{file=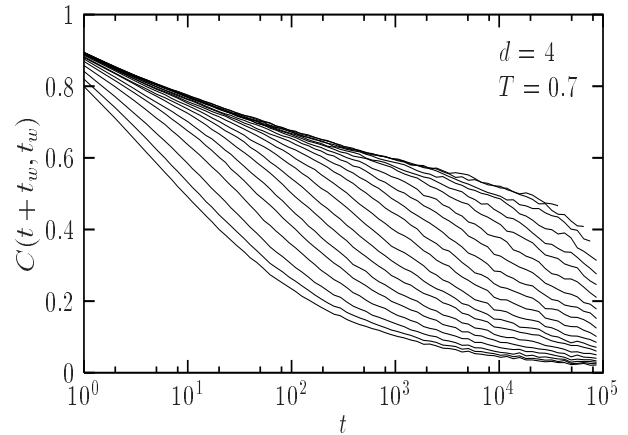,width=8.6cm,height=6.5cm} 
\end{center}
\caption{Two-time autocorrelation functions $C(t_w+t,t_w)$
in simple aging experiments in $d=4$ and $T=0.7$. We show 20 different 
waiting times, which are logarithmically spaced in the 
range $t_w \in [2,57797]$ and increase from left to right.}
\label{isocorr}
\end{figure}

As is now well-documented~\cite{review_aging}, the slow evolution 
of glassy materials following a quench is best analysed through the measurement
of a two-time quantity, typically susceptibility or correlation functions. 
Here, we measure in the process of isothermal
aging the two-time spin-spin correlation function $C(t_w+t,t_w)$ of the 
system.
This quantity can also be accessed experimentally through careful  
noise measurements \cite{Herisson}.
It is represented as a function of the time
difference $t$ in Fig.~\ref{isocorr} for the 4 dimensional sample. 
Similar curves are obtained at all temperatures, in $d=3,4$.
We get the well-known `two-step' decay of the correlation
function with a first, stationary, part followed by a second,
non-equilibrium, aging part. 
The existence of these different regimes is
easily understood qualitatively, but a more quantitative description
of both time sectors is not completely settled yet.

From a theoretical point of view, both mean-field models~\cite{jpa} and the
multi-layer trap model~\cite{multitrap1} 
predict that the short-time and long-time 
contributions are additive,
\begin{equation}
C(t_w+t,t_w) \simeq C_{\rm eq}(t) + C_{\rm aging}(t,t_w),
\label{addition}
\end{equation}
whereas aging at a critical point leads to multiplicative scalings 
\cite{GodrecheLuck}, 
\begin{equation}
C(t_w+t,t_w) \simeq C_{\rm eq}(t)  C_{\rm aging}(t,t_w).
\label{multi}
\end{equation}
as used both 
in simulations~\cite{heiko} and in early analysis of experimental 
data~\cite{review_manip2,manips}.
The equilibrium part can be fitted, both experimentally and numerically, 
by a power-law
\begin{equation}
C_{\rm eq}(t) \simeq A t^{-x(T)},
\label{powerlaw}
\end{equation}
with a temperature dependent exponent $x(T)$, which takes rather small values.
These two forms (additive and multiplicative) 
are actually not very different for short times, since 
$C_{\rm aging}(t,t_w)$ is approximately constant for $t \ll t_w$, in the
regime where $C_{\rm eq}(t)$ varies most.
However, one should stress that the extrapolation of the 
multiplicative scaling behavior (\ref{multi}) 
associated to (\ref{powerlaw})
to large times implies a zero Edwards-Anderson parameter,
defined dynamically as: 
\begin{equation}
q_{\rm EA} = \lim_{t \to \infty} \lim_{t_w \to \infty} C(t_w+t,t_w).
\end{equation}
Indeed, no clear plateau appears in the curves of Fig.~\ref{isocorr}.
On the other hand, the additive scaling suggests a non-zero value of 
$q_{\rm EA}$, and accounts well for 
the experimental data \cite{review_manip2,Herisson}. 

Various scaling forms have been predicted for the 
aging contribution.
In mean-field models, one expects an `ultrametric' behavior~\cite{jpa} 
\begin{equation}
C_{\rm aging}(t,t_w) = \sum_i {\cal C}_i \left( \frac{h_i(t_w+t)}{h_i(t_w)} 
\right),
\end{equation}
where the infinite sum over the 
index `$i$' refers to various `time sectors'~\cite{review_aging,jpa}, and
the various functions ${\cal C}_i$ and
$h_i$ have yet unknown functional forms~\cite{jpa}. 
An explicit example of such a scaling has recently been given in
Ref.~\cite{Bertin},
in the context of the trap model, where the infinite sum boils down to
\begin{equation}\label{logs}
C_{\rm aging}(t,t_w) =  {\cal C}\left(\frac{\ln t}{\ln t_w}\right).
\end{equation}
Note that it is $\ln t$ and not $\ln (t_w+ t)$ that appears in this
equation,
which ensures dynamic ultrametricity~\cite{Bertin,BBKultra}. This scaling 
is not observed experimentally, except perhaps for $t \ll t_w$ (see 
Ref.~\cite{Bertin}).
The scaling (\ref{logs}) is similar to, but different from, the scaling 
form
suggested by the droplet picture, where~\cite{drop,FLDM}, 
\begin{equation}
C_{\rm aging} (t,t_w) \simeq {\cal C} \left( \frac{\ell_T(t+t_w)}{\ell_T(t_w)} 
\right), 
\end{equation}
with $\ell_T(t_w) \sim (\ln t_w)^{1/ \psi}$~\cite{drop}. 
The `droplet' scaling variable $\ln (t+t_w) / \ln t_w$ suggests 
super-aging, i.e. an effective 
relaxation time growing faster than the age of the system $t_w$, which 
is not borne out by 
experimental data showing instead 
a tendency towards sub-aging. We come back to this point below.

In the absence of any compelling theoretical description,
both experiments and simulations have been 
{\it phenomenologically} fitted with some scaling functions of the type
\begin{equation}
C_{\rm aging}(t,t_w) \simeq {\cal C} \left( \frac{h(t+t_w)}{h(t_w)} \right),
\label{functionh}
\end{equation} 
where the function $h(u)$ is given various functional forms 
\cite{review_aging}, 
related to the often debated~\cite{review_manip2,federico,BEPJB}
issue of sub-aging versus super-aging behavior.
A widely used form for $h(u)$, which we adopt here, 
is \cite{review_manip2,struik} 
$h(u) = \exp \left[ u^{1-\mu}/(1-\mu) \right]$, where
the exponent $\mu$ allows one to interpolate between super-aging ($\mu > 1$)
and sub-aging ($\mu < 1$), via simple aging ($\mu=1$, for 
which $h(u)=u$). The effective
relaxation time is indeed given by $t_{\rm rel} \sim t_w^\mu$.
Note that if one takes $h(u) = \ell_T(u)$ with $\ell_T$ 
given by Eq.~(\ref{ellt}), then, as in the droplet model, super-aging would
also be observed for long waiting times, 
since the effective relaxation time, defined as $h(t_w)/h'(t_w)$, 
now grows with the waiting time as
\begin{equation}
t_{\rm rel} (t_w)\sim t_w \left( z_c +
\frac{\Upsilon(T) \psi \ell_T^\psi}{T} \right) \gg t_w, \,\,\,
\ell_T(t_w) \gg \xi(T).
\label{super}
\end{equation}
Note finally that the existence of a growing coherence length 
$\ell_T(t_w)$ in spin-glasses
does not necessarily implies that the correlation function can be expressed 
in terms 
of this length scale only. Indeed, since time scales are broadly distributed, 
processes corresponding to different lengths scales are expected to mix 
together. This may also happen in simpler models~\cite{BEPJB}.

\subsection{Numerical results}

\begin{figure}
\begin{center}
\psfig{file=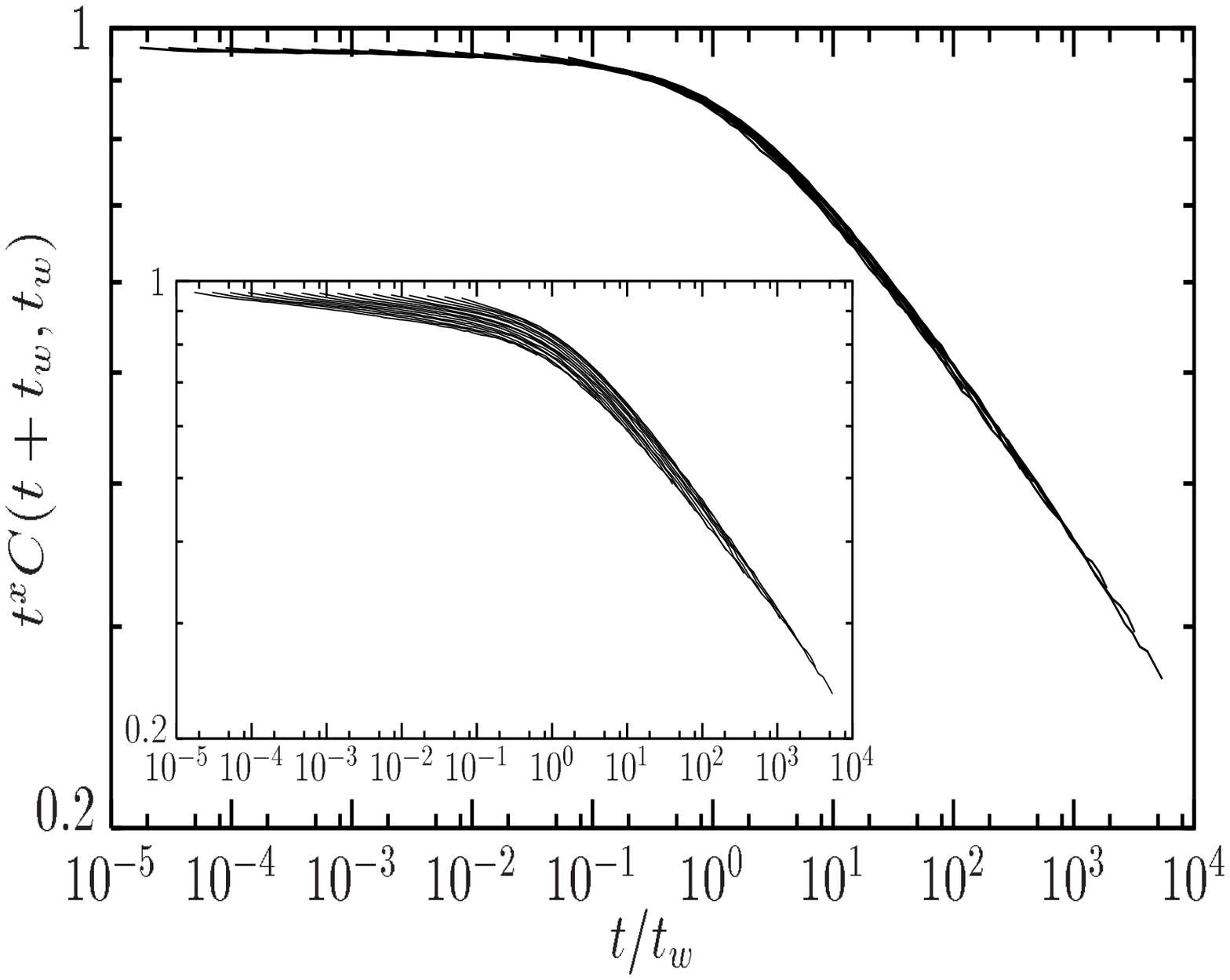,width=8.6cm,height=6.5cm} 
\psfig{file=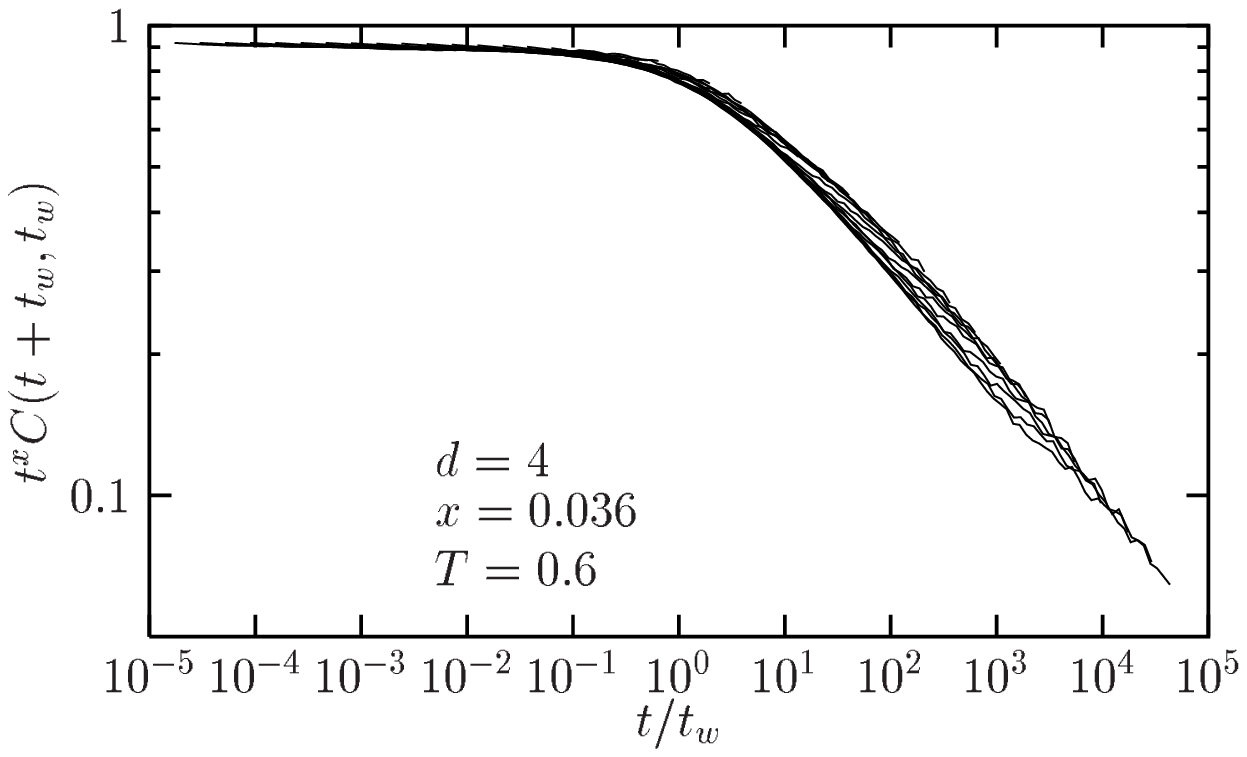,width=8.6cm,height=6.5cm}
\end{center}
\caption{Rescaled autocorrelation functions according to the 
scaling form~(\ref{multi}) in $d=3$ (top, $T=0.6$, $x=0.0128$) 
and $d=4$ (bottom).
The inset of the top figure is the simpler scaling form 
$C(t+t_w,t_w) = {\cal C}(t/t_w)$.}
\label{isocorrresc}
\end{figure}

We now discuss our numerical results.
We first show in the inset of Fig.~\ref{isocorrresc} (top)
that the simple scaling $C(t+t_w,t_w) \sim {\cal C} (t/t_w)$ obviously 
fails in describing the data. Neither short nor long time scales
are correctly described by this form.

We then show in Fig.~\ref{isocorrresc}
that when the short-time dynamics is taken into account through
the multiplicative scaling form~(\ref{multi}), the collapse looks
almost perfect in $d=3$ (Fig.~\ref{isocorrresc}, top), whereas a small 
{\it super-aging} trend subsists in $d=4$ (Fig.~\ref{isocorrresc}, bottom).
Indeed, in this $t/t_w$ representation, older curves are still above
the younger ones, suggesting that rescaling the time by $t_w$ 
is not sufficient to superimpose all the curves.
Hence, the introduction of another fitting parameter is required 
to describe the data, namely the exponent $\mu$ defined above. 

\begin{figure}
\begin{center}
\psfig{file=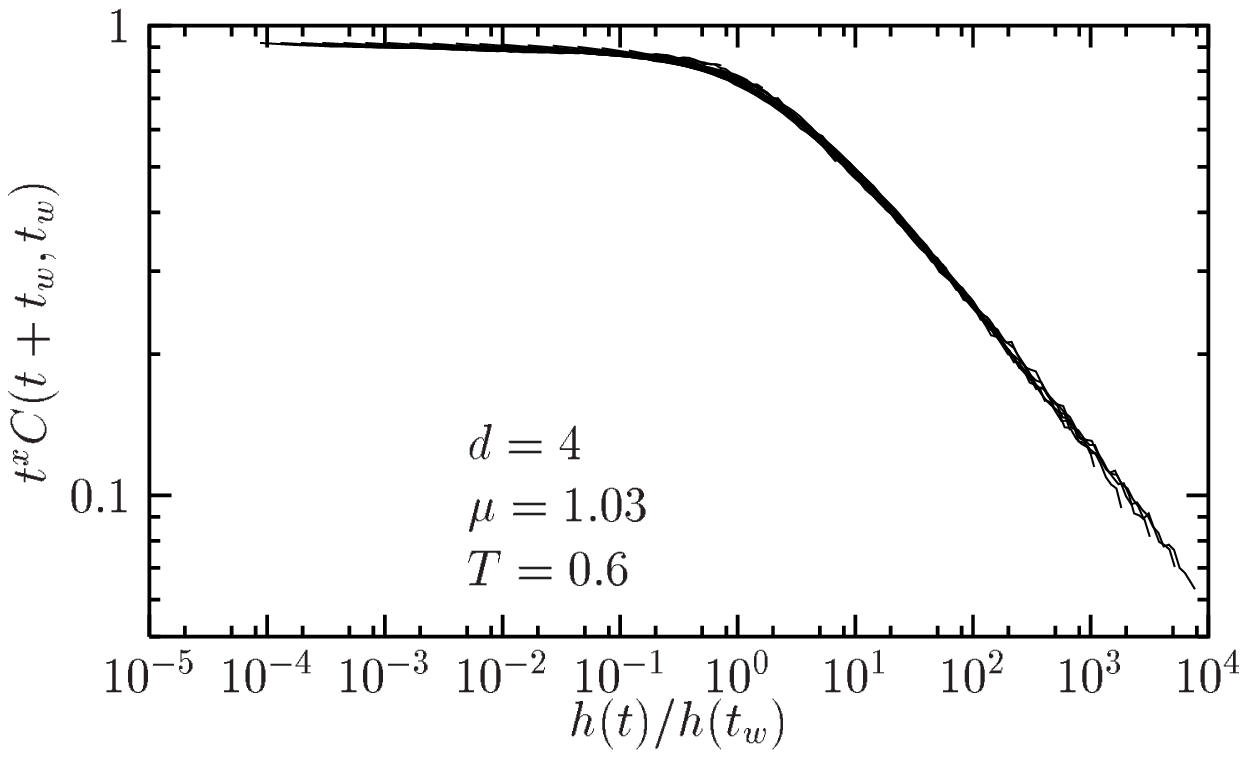,width=8.6cm,height=6.5cm} 
\psfig{file=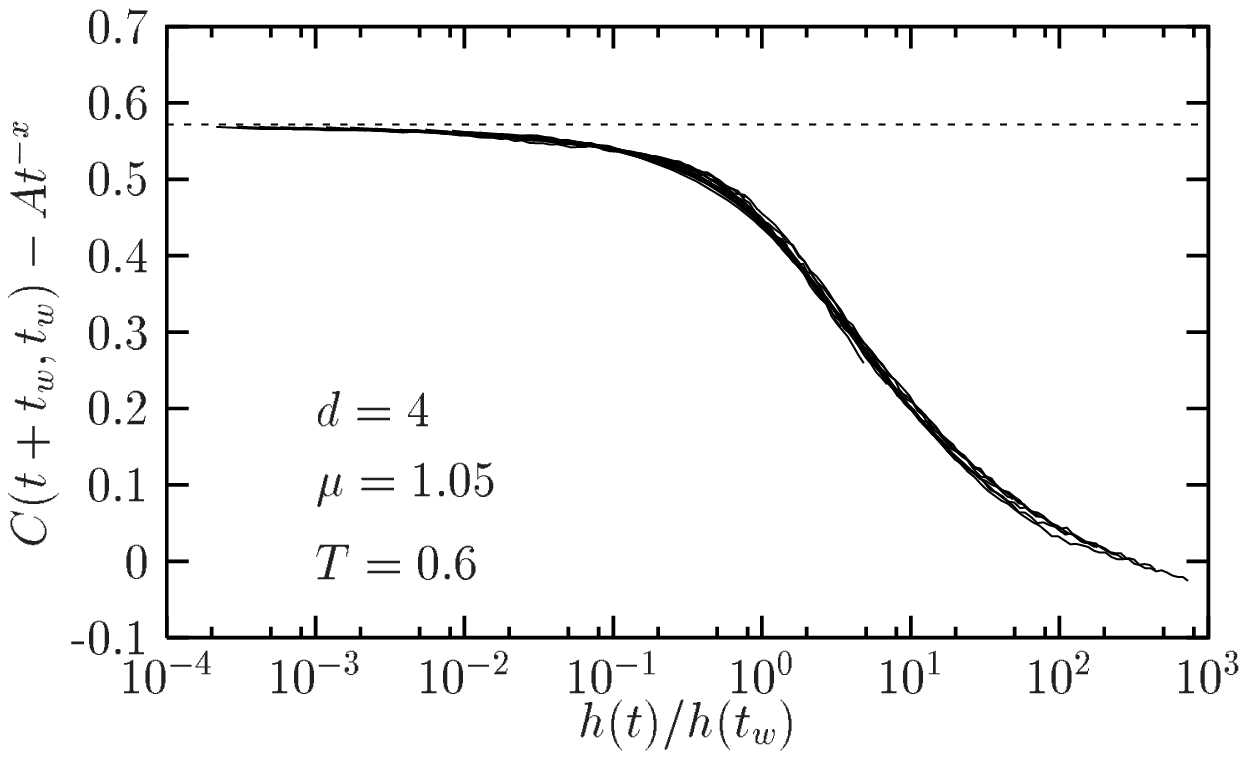,width=8.6cm,height=6.5cm} 
\end{center}
\caption{Top: Rescaled autocorrelation functions
according to the scaling forms (\ref{multi}) and (\ref{functionh}), thus
allowing for $\mu \ne 1$.
Bottom: Rescaled autocorrelation functions according to 
the scaling forms (\ref{addition}) and (\ref{functionh}),
{\it choosing} parameters so that
data extrapolate to a non-zero 
Edwards-Anderson parameter, as shown by the horizontal 
dashed line.}
\label{isocorrresc2}
\end{figure}

We find that with $(x,\mu)$ as free parameters
and the multiplicative 
form (\ref{multi}), the data can be nicely collapsed 
for the {\it whole} temperature range studied, $0.4 \le T \le 1.0$, 
in both dimensions $d=3$ and $d=4$.
An example of such a rescaling is given
in Fig.~\ref{isocorrresc2} (top).
In $d=3$, our data are consistent with $\mu = 1$ in the whole temperature 
range, and we find an exponent $x(T)$ in close agreement with values reported
in Ref.~\cite{heiko}.
In $d=4$, our finding for $x(T)$ also follow the reported 
values~\cite{Ricci96}.
In addition, as suggested by Fig.~\ref{isocorrresc}, 
we find that the exponent $\mu$ has to be {\it larger} than 1
for $T \le 0.6$, while $\mu = 1$ is compatible with the data for 
$0.8 \le T \le 1.0$.
This observation was never reported, although 
a re-analysis of published data~\cite{Ricci96} 
confirms this trend~\cite{Ricci2}.
In both dimensions, we find that the scaling 
function ${\cal C}(x)$ behaves 
as ${\cal C}(x) \simeq const$ when $x \ll 1$, and as
${\cal C}(x) \simeq x^{-\lambda}$ for $x \gg 1$, as in Ref.~\cite{heiko}.

Close to $T_c$ (i.e. $T=1.0$), we find $\mu=1$.
This is physically expected, since standard 
non-equilibrium critical dynamics gives indeed 
the scaling (\ref{multi}), with $h(t) \sim \ell_T(t) \sim t^{1/z_c}$.
In that case, the coherence length is the usual dynamic correlation 
length~\cite{GodrecheLuck,janssen,behose}.
The fact that the scaling function $h(u)$ changes when the temperature
is lowered in $d=4$ suggests that the dynamics leaves the critical regime. 
It is thus a priori surprising that the multiplicative power-law scaling
still holds for low temperatures. 
The situation appears different in $d=3$, where the dynamics does not show any clear 
change when the temperature is lowered below $T_c$. 

One can therefore try to
rescale the $4d$ data according to an additive scaling, which allows for
a non-zero Edwards-Anderson parameter, Eq.~(\ref{addition}).
In this case, we have three free parameters, $(x, \mu, A)$, where 
$A$ is the amplitude of the stationary part $A$. This is unfortunately 
too much since in this case $A$ (and thus $q_{\rm EA}$) is very poorly
constrained.
As noted in Ref.~\cite{federico}, the values of the parameters
$A$ and $x(T)$ are in fact 
strongly anti-correlated, but the data is insufficient to
pin down their individual values, and hence 
to conclude on the value of $q_{\rm EA}$.
However, noting that $C_{\rm aging}(x) > 0,  \forall x$ allows us 
to give a possible range for $q_{\rm EA}$.
For instance, for $d=4$ and $T=0.6$, we find 
that $q_{\rm EA} \in [0.57, 0.68]$ leads to a reasonably good 
rescaling of the data. These values are in agreement with previous estimations
of the Edwards-Anderson parameter~\cite{Ricci96}.

Interestingly, the additive procedure has 
little impact on the value of the aging exponent $\mu$.
For $d=3$, we still find that $\mu =1$ allows for a good rescaling of 
the data, whereas in $d=4$ 
the super-aging tendency is slightly reinforced by this rescaling. An example 
of this is shown in Fig.~\ref{isocorrresc2} (bottom).

\subsection{Conclusion}

From the above analysis of our numerical results on isothermal aging, we 
conclude on the following.

(i) Although data are compatible with the standard scenario 
where $q_{\rm EA} > 0$ for 
$T < T_c$, we cannot rule out, from our numerical study of 
spin-spin correlation function, the fact that asymptotically 
$q_{\rm EA}=0$ both in $d=3$ and in $d=4$. 
Longer simulations in $d=4$ \cite{Ricci96} however seem to favor the 
additive scaling over the multiplicative
scaling, and therefore a non-zero Edwards-Anderson parameter.

(ii)
The exponent $x$ that describes the short-time decay of the correlation
function also describes the behavior of the equilibrium a.c. susceptibility.
The measurement of the latter allows then to obtain the value
of $x$ independently. 
It was then shown experimentally that when this is done, 
the additive scaling (\ref{addition}) works well~\cite{review_manip2,manips}. 

(iii) Whatever the chosen rescaling for the short-time dynamics, 
we find a systematic super-aging behavior in $d=4$. This is consistent with 
the identification of the scaling function $h(u)$ with a coherence length
$\ell_T(u)$, growing like Eq.~(\ref{ellt}).
Indeed, the resulting relaxation
time given by Eq.~(\ref{super}) can be written approximately as $t_w^\mu$ with
\begin{equation}
\mu-1 = \frac{\psi}{z_c}.
\end{equation} 
On the other hand, no super-aging is found in $d=3$, although, as shown 
below, Eq.~(\ref{ellt}) also seems to hold. However
the distinction between Eq.~(\ref{ellt}) and a pure power-law with a 
temperature dependent exponent will be much harder to establish in $d=3$.

(iv) Our data show no tendency towards sub-aging, 
in contrast to what is consistently found in all 
experiments~\cite{review_manip2}. We shall see below that the introduction 
of a finite cooling rate (instead of the direct $T=\infty \to T$ 
quench considered
in this section) in fact results in an effective sub-aging. 

\section{Growth of a coherence length}
\label{growthsec}

We now turn to a more {\it geometric} characterisation of aging, and try to
associate the stationary part of the correlation function to 
equilibrated small scale dynamics, 
and the aging part of the correlation function 
to out-of-equilibrium, large scale dynamics. 
The time dependent crossover scale is a {\it coherence
length}, that would be the domain size in a coarsening
ferromagnet~\cite{bray}, or the dynamic 
correlation length at the critical point~\cite{janssen}. 
In the case of spin-glasses, a more subtle definition is 
needed~\cite{huse,heiko}.
As for the autocorrelation function, 
we discuss in detail the physics involved in the scaling 
form of this function.
Moreover, we extend previous works in $d=4$
to a larger temperature range. This is necessary
in order to extract the parameters involved in Eq.~(\ref{ellt}).
The latter analysis is also new in $d=3$, and allows a direct comparison
with experiments.

\subsection{The four-point correlation function}

\subsubsection{Definition}

\begin{figure*}
\begin{center}
\psfig{file=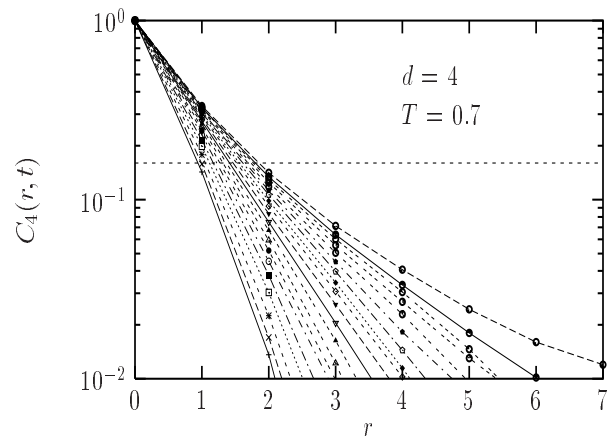,width=8.6cm,height=6.5cm} 
\end{center}
\caption{4-point correlation function for $T=0.7$ in $d=4$
measured in an isothermal aging experiment.
Times are logarithmically spaced in the range $t_w \in [2,57797]$ and
increase from left to right. The horizontal dotted line corresponds to
$q_{\rm EA}^2$, using $q_{\rm EA}=0.4$ from static studies.}
\label{4point}
\end{figure*}

As proposed by several authors~\cite{huse,heiko}, the coherence length can 
be measured in a simple isothermal aging protocol from the spatial structure
of the 4-point correlation function $C_4(r,t)$, defined 
in Eq.~(\ref{c4r}).
This 4-point function is the analog of the structure factor 
in a usual domain growth problem~\cite{bray}, adapted to the case
of the disordered system under study, where any growing pattern is random
and can only be identified by comparing two independent real replicas of 
the same system, prepared at $t=0$ in a different random state. 
It measures the similarity of the relative spin orientations in the 
two systems at a distance $r$ after time $t$. 
Typical results are presented in Fig.~\ref{4point}.
The spatial decay of this correlation function becomes slower when
the time increases, clearly indicating the growth of a length scale
in the system. This was already noted several 
times~\cite{hajime,huse,heiko,Marinari}.

\subsubsection{Functional form of $C_4$}

The correct identification of the coherence length $\ell_T(t_w)$ is however
not completely straightforward. Indeed, the naive definition 
\begin{equation}
C_4 \big(r=\ell_T(t_w),t_w \big) = c,
\label{defell}
\end{equation}
where $c$ is an arbitrary constant, say $c=0.1$, leads to 
inconsistent results, because
the decay of $C_4$ is not purely (or even possibly stretched) 
exponential~\cite{comment}.
This fact is very clear in $d=4$, where the definition (\ref{defell})
with $c = 0.1$ leads to a coherence length which is such that for some 
$t_w$, $\ell_{T_1}(t_w) > \ell_{T_2}(t_w)$ when $T_1 < T_2$, i.e. a 
faster growth  
at lower temperatures. This result is physically unacceptable.

From a more careful analysis of the data~\cite{comment}, 
one finds that $C_4$ receives two contributions:
a `quasi-equilibrated' decay for $r < \ell_T(t_w)$, followed by a 
`non-equilibrium' decay at large distances. This reflects the fact
that at time $t_w$, 
the system has equilibrated up to a length scale $\ell_T(t_w)$, 
with a non trivial equilibrium correlation function. 
This is the direct analog of the `two-step' behavior observed in 
the autocorrelation function. 

As suggested by previous studies~\cite{Marinari,comment}, a possible
functional form is
\begin{equation}\label{c4}
C_4(r,t_w) =  \frac{1}{r^{\alpha(T)}}\, {\cal C}_4 \left( 
\frac{r}{\ell_T(t_w)} \right),
\end{equation}
with a temperature dependent $\alpha(T)$, and ${\cal C}_4(x)$ a scaling 
function.
It is difficult to confirm or dismiss this result, since 
the numerical correlation functions typically decay over $3 - 5$ 
lattice spacings only, and other functional forms are possible (see below).
Note that very few equilibrium data is available for this 
correlation function~\cite{MARIRI}, which would be a very interesting
information to compare with Eq.~(\ref{c4}).

As noted in Ref.~\cite{comment}, Eq.~(\ref{c4}) suggests that 
$C_4(r \to \infty,t_w=\infty)$ tends to zero.
This must be
contrasted with the prediction of
the droplet picture or any other theory in which the overlap distribution
is a trivial $\delta$-function at $q_{\rm EA}$, where~\cite{drop}
\begin{equation}\label{dropdrop}
\lim_{r \to \infty} \lim_{t_w \to \infty}  
C_4(r,t_w) =  q_{\rm EA}^2 + O(r^{-\theta}),
\end{equation}
where $q_{\rm EA}$ is the 
Edwards-Anderson parameter and $\theta$ the energy exponent, estimated to
be $\sim 0.2$ in $d=3$~\cite{BM} 
and $\sim 0.7$ in $d=4$~\cite{theta4}, i.e., smaller than the
values of $\alpha$ reported in Table~\ref{tablealpha} below.

Although Fig.~\ref{4point} suggests that $C_4(r,t_w)$ is rapidly
much smaller than $q_{\rm EA}^2$, and compatible with $C_4(r \to \infty,\infty)=0$,
we find that it is still possible to rescale 
the data according to scaling
forms which imply a non-vanishing large distance limit.
In order to illustrate this point, we tried 
the following ansatz (not motivated by any theoretical argument):
\begin{equation}\label{c4bis}
C_4(r,t_w) =  \big( q_{\rm EA}^2 + a \exp(-br) \big)\, 
{\cal C}'_4 \left(\frac{r}{\ell_T(t_w)} \right),
\end{equation}
which allows a rescaling of the data as good as Eq.~(\ref{c4}).
In this case, the stationary part of $C_4$ indeed 
tends towards $q_{\rm EA}^2$,
but much {\it faster} than $r^{-\theta}$. More generally, the inequality 
$\alpha(T) > \theta$ makes Eq.~(\ref{dropdrop}) rather unplausible. 

\subsubsection{Discussion}

A word of caution is however needed here. Although a non-zero value of the 
Edwards-Anderson parameter $q_{\rm EA}$ is expected in the spin-glass phase, 
dynamical evidence for this is still rather weak, at least in
$d=3$. For example, as discussed
in the previous section, the dynamic spin-spin correlation function 
cannot rule out a non-zero value of $q_{\rm EA}$. 
It is well-known that on the time scale (resp.
system size) of dynamic (static) simulations, the apparent value of
$q_{\rm EA}$ constantly shifts towards $0$ with increasing times or 
sizes~\cite{review_simu}.
In that sense, the evidence that $C_4(r,t_w)$ tends to zero 
at large distances could be compatible with a 
very small value of $q_{\rm EA}$. The evidence against the simplest 
droplet picture
is however stronger in $d=4$, since the numerical evidence for $q_{\rm EA}> 0$
is more compelling in this case.

An alternative interpretation in three dimensions 
is that $q_{\rm EA} = 0$ at all temperatures, which means 
that either there is no true spin glass transition, or else that the nature 
of the spin glass phase is different from what has insofar been theoretically
expected.
This issue might also be related to the 
existence of large excitations of finite energy recently 
found in Ref.~\cite{lamarcq}. 
For instance, one could be in a Kosterlitz-Thouless ({\sc kt}) like 
situation where $q_{\rm EA}=0$,
but the 4-point correlation function changes from exponential for 
$T>T_c$ to power-law for $T<T_c$ with, possibly, a temperature 
dependent exponent $\alpha(T)$. Along this (speculative) line of thought, it has
been pointed out 
recently that the dynamics of a critical {\it phase} 
(such as the {\sc kt} phase) 
shares many similarities with spin glass dynamics \cite{BerthierEPL,behose}.

\subsection{Numerical results for the coherence length}

We thus adopt a phenomenological definition of $\ell_T(t_w)$ as the 
time-dependent length which leads, using Eq.~(\ref{c4}), to the best numerical 
collapse  of $C_4(r,t_w)$ measured at different times. 

It is important to note that the 
numerical value of the exponent $\alpha(T)$ used in this scaling plot has a 
significant influence on the resulting 
growth law for $\ell_T(t_w)$. Since the spatial support of $C_4$ is very small,
it is impossible to determine this exponent numerically with great 
accuracy.
The conclusion is that even using the above scaling procedure, there 
is still some
degree of arbitrariness in the definition of the coherence length $\ell_T$.

We first report 
in Table~\ref{tablealpha} the values of $\alpha(T)$ 
found in our simulation  
for the $d=3$ and $d=4$ cases.
Our values for $d=3$ are quite close to the ones found 
in Ref.~\cite{Marinari}. For $d=4$, only the value of 
$\alpha(T=1.0) = 1.63$ has been 
reported in Ref.~\cite{Ricci96}, in excellent agreement with our 
determination.

\begin{figure}
\begin{center}
\psfig{file=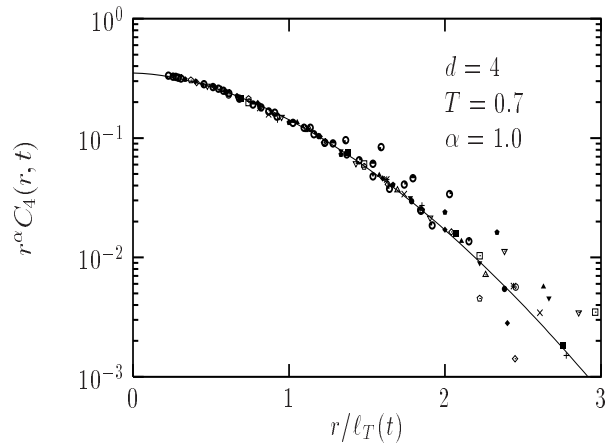,width=8.6cm,height=6.5cm} 
\end{center}
\caption{Rescaled 4-point correlation function for $T=0.7$ in 
$d=4$. The full line is a fit of the scaling function 
${\cal C}_4(x)$ to a `stretched' exponential form,
with $\beta=1.7$.}
\label{4pointresc}
\end{figure}

An example of data collapse for the 4-point correlation function can be 
seen in 
Fig.~\ref{4pointresc}. As in Ref.~\cite{Marinari}, we find that the cut-off 
function ${\cal C}_4$ 
is compatible with a `stretched' exponential
form, ${\cal C}_4(x) \sim 
\exp(-x^\beta)$, but with an exponent $\beta > 1$.

\begin{table}[t]
\centering
\begin{tabular}{c|c|c}
$T$ & $\alpha(T)$ in $d=3$ & $\alpha(T)$ in $d=4$ \\ \hline
1.0 & 0.6  & 1.63 \\
0.9 & 0.5  & 1.35\\
0.8 & 0.5  & 1.25\\
0.7 & 0.5  & 1.0\\
0.6 & 0.5  & 0.9\\
0.5 & 0.45 & 0.9
\end{tabular}
\caption{Values of the exponent $\alpha(T)$ in $d=3$ and $d=4$. 
Note that $\alpha(T)$
is nearly constant in $d=3$ but significantly changes with 
temperature for $d=4$. 
This will turn out to be important in the following.}
\label{tablealpha}
\end{table}

\begin{figure}
\begin{center}
\psfig{file=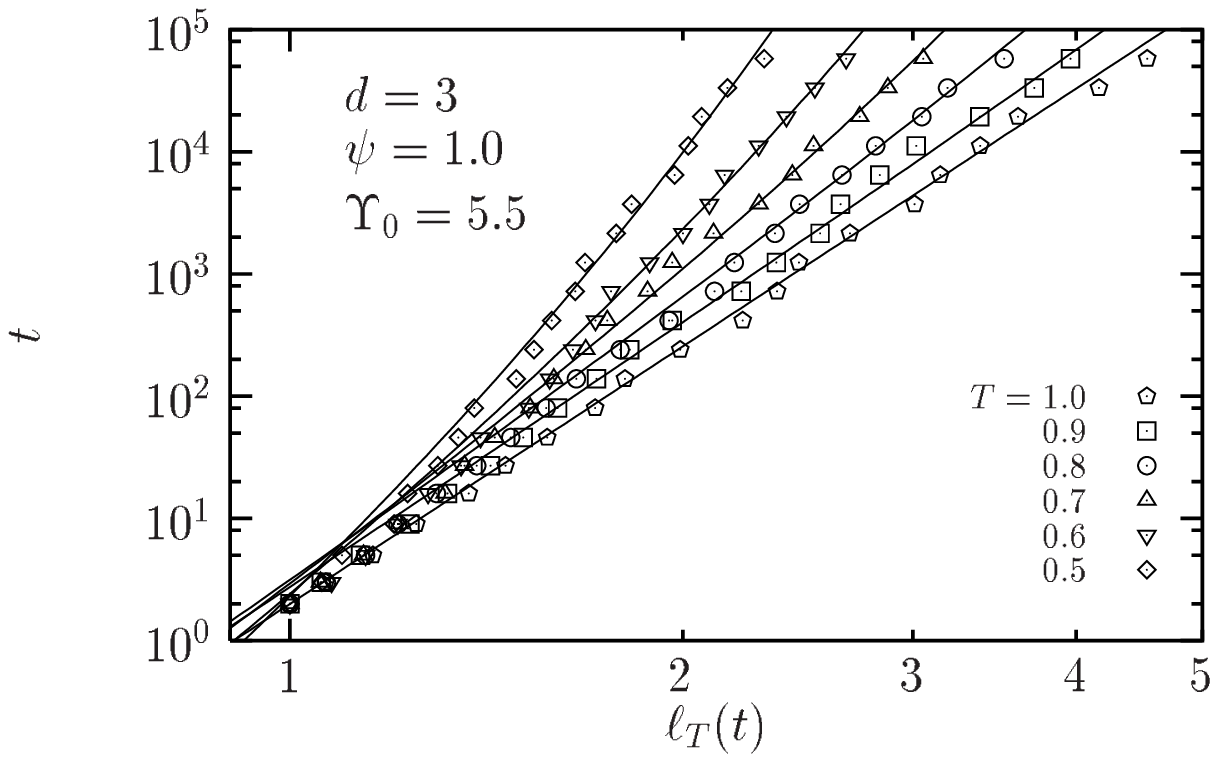,width=8.6cm,height=6.5cm} 
\psfig{file=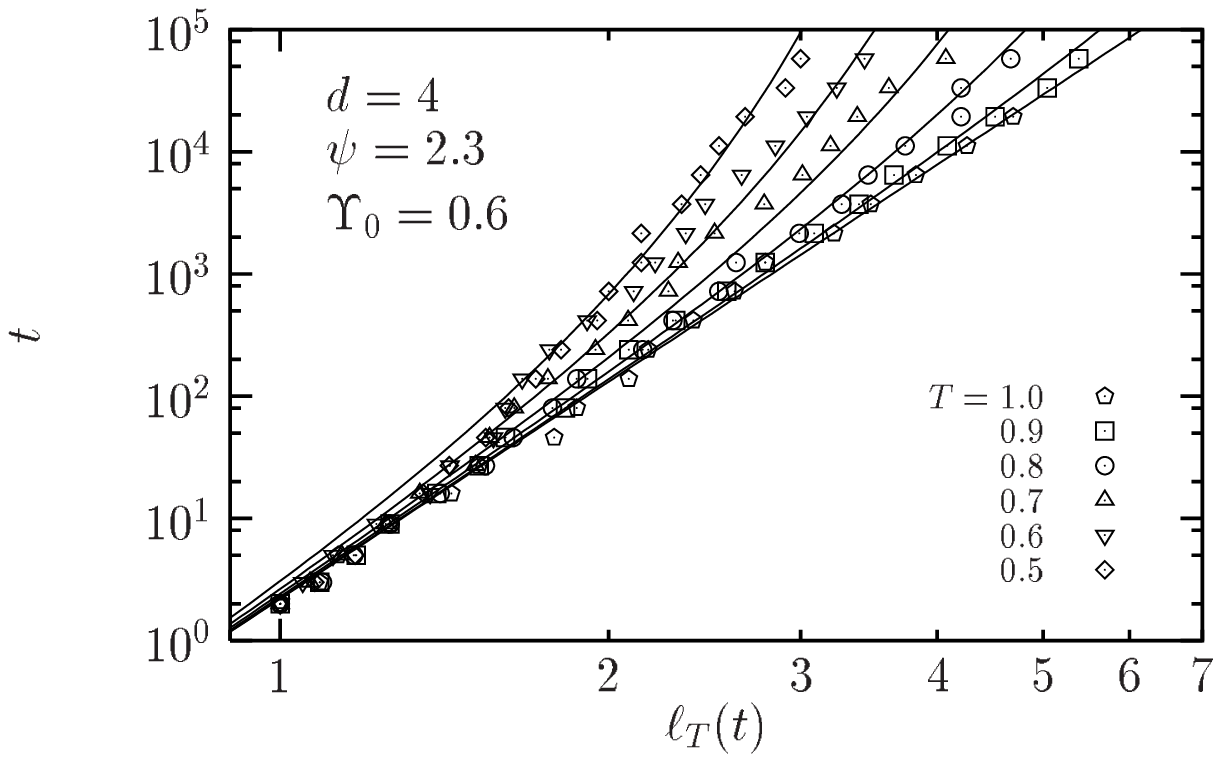,width=8.6cm,height=6.5cm} 
\end{center}
\caption{Growth laws of the coherence length 
in 3 dimensions (top) and 4 dimensions (bottom).
The points are the data, the full lines are fits to Eq.~(\ref{ellt}).}
\label{JP2}
\end{figure}

The growth of $\ell_T(t_w)$ for different temperatures, $0.5 \le T \le 1.0$,
is reported in $d=3$ and $d=4$ in Fig.~\ref{JP2}.
From Eq.~(\ref{ellt}), we expect the coherence length to grow as 
a power law at short times. This is the critical regime
characterized by $\ell_T(t_w) \ll \xi(T)$
such that
the growth law is $\ell_T(t_w) \sim t_w^{1/z_c}$. 
For larger times, the exponential activated term 
in Eq.~(\ref{ellt}) should slow 
down the growth of $\ell_T(t)$ in a temperature dependent manner.
The locus of the crossover itself must be temperature dependent. 
All this is observed numerically, see Fig.~\ref{JP2}.
This is also qualitatively consistent with the law 
extracted from experiments~\cite{JPB,Dupuis} 
and reported in Fig.~\ref{growth}.

We are now in position to compare 
the growth law obtained numerically to Eq.~(\ref{ellt}).
The critical exponent $\nu$ is taken from previous numerical work. 
We take $\nu=1.65$ in $d=3$ and $\nu=0.8$ in 
$d=4$~\cite{review_simu}.
The exponent $z_c$ and the microscopic time $\tau_0$
are fixed by the data at $T=T_c$.
We find $z_c \sim 7.0$ and $\tau_0 \sim 2.0$ in $d=3$,  
$z_c \sim 5.9$ and $\tau_0 \sim 2.2$ in $d=4$. 
The values for the dynamic exponents are compatible with 
previous determinations~\cite{Marinari,Ricci96}.

We are thus left with $\psi$ and $\Upsilon_0$ as free parameters.
We find that Eq.~(\ref{ellt}) accounts very well for the data in $d=4$ 
with $\psi \sim 2.3$ and $\Upsilon_0 \sim 0.6$, see Fig.~\ref{JP2}.
In 3 dimensions, we were not able to use Eq.~(\ref{ellt}) with 
a fixed $\tau_0$.
Instead, the
fits reported in Fig.~\ref{JP2} give $\psi \sim 1.0$ and $\Upsilon_0 
\sim 5.5$,
but were obtained by letting $\tau_0$ to be temperature dependent, with a 
non-monotonic temperature behaviour, for which there is 
of course no physical explanation.
Simpler power law fits with a temperature 
dependent $z(T)=z_c T_c/T$ but a constant $\tau_0=1$ give equally good 
results, with much less free parameters.
This might indicate:
\begin{itemize}
\item either that, as discussed above, 
the whole Ising spin-glass phase in three dimensions is 
Kosterlitz-Thouless like, where the dynamics is indeed described 
by power-laws for all temperatures. 
This would be compatible with the fact that no cross-over beyond the critical
regime is detected in the evolution of
$C(t,t_w)$ and $C_4(r,t_w)$ in $d=3$;

\item or that Eq.~(\ref{ellt}) is too simple to quantitatively reproduce 
the detailed crossover between critical and activated dynamics in $d=3$. 
It must
be noted that, as emphasized in Refs.~\cite{BerthierEPL,mikemoore}, 
the simulations
are performed right in the regime where the influence of the critical point
is still strong;

\item or else that the procedure to extract $\ell_T(t_w)$ from $C_4(r,t_w)$ is
somewhat biased.

\end{itemize}

We should nevertheless add the following
remarks about the three dimensional case.

(i) As will be clear below, the simple power-law
growth of $\ell_T$ with $z(T)=z_c T_c/T$ cannot explain the small 
temperature shift 
effects, that suggest -- both numerically and experimentally --
deviations from a pure activated growth.

(ii) One can also 
extract from the data 
the local slope of $\log t(\ell_T)$ as a function of $\log \ell_T$, 
which should be independent 
of $\ell_T$ for a pure power law. One finds instead 
systematic deviations, such that the {\it effective exponent $z$ 
indeed increases} with $\ell_T$, as predicted by Eq.~(\ref{ellt}).
Moreover, the amplitude of these deviations
vanish as $T$ increases towards $T_c$, in a way 
very much compatible with  Eq.~(\ref{ellt}).

(iii) Finally, it has been suggested in the past that a pure power-law 
behaviour for $\ell_T(t_w)$ is associated with `Replica Symmetry Breaking',
which predicts that the whole low temperature phase is `critical'. 
We disagree 
with this point of view: the growth of $\ell_T(t_w)$ could be 
asymptotically logarithmic, 
as in the droplet picture, even if the equilibrium phase is not unique. 
This seems 
to be two totally separate issues as long as one does not associate 
$\ell_T$ with the
size of compact droplets.

\section{Probing the barriers: Small temperature-shift experiments}
\label{Tshifts}

In order to directly probe the influence of the barriers on the aging
dynamics, we perform
numerically the analog of temperature-shift experiments~\cite{shift1}.
Similar simulations were already performed
in Refs.~\cite{hajime,heiko2} with qualitative results only.
Here we go much further and perform, 
as was done in Ref.~\cite{JPB}, a detailed quantitative
analysis of the results.
The protocol is the following. 
The system is quenched from $T=\infty$ to the temperature
$T_1=T_2 \pm \Delta T$ where it ages; $\Delta T$ here will always be positive.
At time $t_w$, the temperature is shifted to $T_2$, where
the measurements start. 
We shall discuss  the 
behavior of the autocorrelation function $C(t_w+t,t_w)$.

\subsection{Aging is less efficient at lower temperatures}

\begin{figure}
\begin{center}
\psfig{file=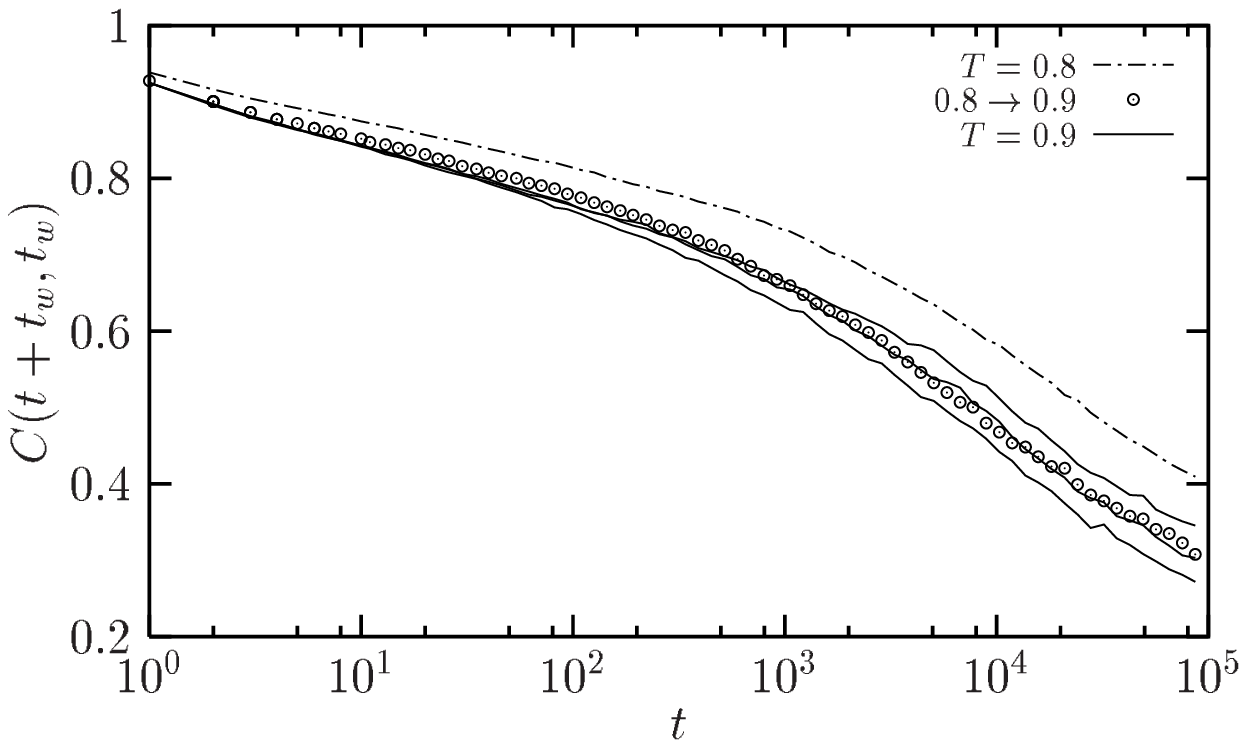,width=8.6cm,height=6.5cm} 
\psfig{file=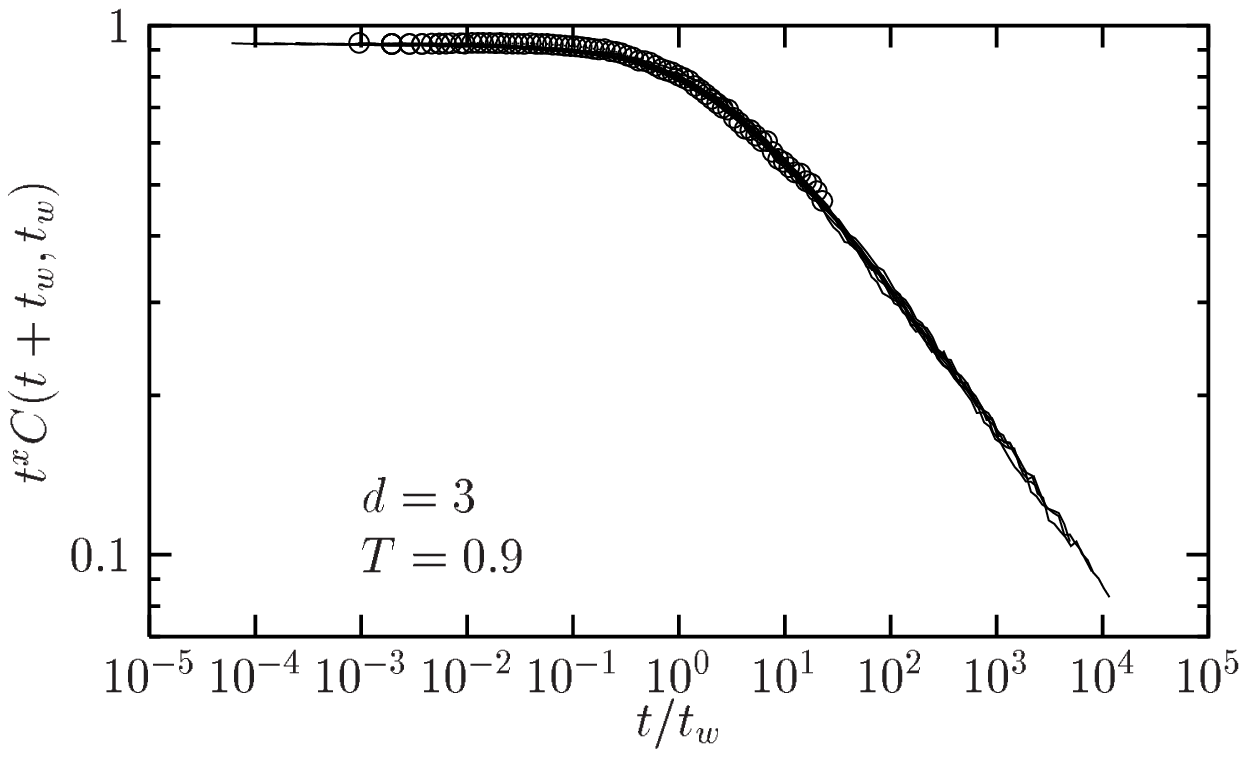,width=8.6cm,height=6.5cm} 
\end{center}
\caption{Comparison between simple aging and shift experiments
in $d=3$.
Top: Lines are simple aging curves, $t_w=3728$ at $T=0.8$ and 
$t_w=1245$, 2154, and 3728 at $T=0.9$ (from bottom to top).
Circles are obtained in the shift 
$T_1=T_2 - \Delta T = 0.8 \to T_2 =0.9$ at time $t_w=3728$. One concludes that 
$1245 < t_w^{\rm eff}  < 3728$. 
Bottom: The correlation obtained in the shift is superposed 
to the correlation obtained in isothermal aging at $T_2$, 
which gives the value $t_w^{\rm eff}=2200$.} 
\label{effective1}
\end{figure}

Let us start with the phenomenology.
We observe that the decay of the correlation
function after the temperature-shift from $T_1=T_2 + \Delta T$ to $T_2$ 
is slower than a purely isothermal aging at $T_2$, 
meaning that preliminary aging at a 
slightly higher temperature `helps' the system at the final temperature.
The opposite effect is found when $T_1=T_2 - \Delta T$, see 
Fig.~\ref{effective1} (top).
In that sense, aging is less efficient at lower temperatures.

Moreover, the decay following the shift
has the same functional form, {\it for small
enough} $\Delta T$, 
as in a simple aging experiment at temperature 
$T_2$. An example of this feature is shown in Fig.~\ref{effective1}.
This implies that the correlation function after the shift
can be superposed 
to the correlation obtained in isothermal aging at $T_2$ by
introducing an {\it effective waiting} time.
One has $t_w^{\rm eff} < t_w$ for $T_2 - \Delta T < T_2$. 
The same effect is observed experimentally
when $\Delta T$ is sufficiently small~\cite{shift1}.

The determination of the 
effective age of the sample can be made rather precise when the 
results of section~\ref{isothermal} are used.
The correlation function obtained in the shift experiment for different 
$\Delta T$ can be 
collapsed on the master curves of Fig.~\ref{isocorrresc2}, using
$t_w^{\rm eff}$ as a {\it single} adjustable parameter. This leads to 
a very precise determination of $t_w^{\rm eff}$, which does not
require any analytical fit of the data, see Fig.~\ref{effective1}.

\subsection{`Time is length': Link with the coherence length}

The effective age of the sample may be simply interpreted 
in terms of length scales. 
The growth of the coherence length being slower at lower temperatures,
one has $\ell_{T_2-\Delta T}(t_w) < \ell_{T_2}(t_w)$.
If one assumes that the age of the sample is fully encoded in the value
of $\ell_T$, then the effective age can be determined by the relation
\begin{equation}
\ell_{T_2-\Delta T}(t_w) = \ell_{T_2} (t_w^{\rm eff}).
\label{estime}
\end{equation} 
This relation will be correct if $\Delta T$ is not too large, such that 
quasi-equilibrated structures of sizes $\lesssim 
\ell_{T_2 - \Delta T}(t_w)$ are almost unchanged by the temperature shift.
In this case, aging at $T_2$ is a simple continuation of aging at 
$T_2-\Delta T$,
at a slightly different rate given by Eq.~(\ref{ellt}).

\begin{figure}
\begin{center}
\psfig{file=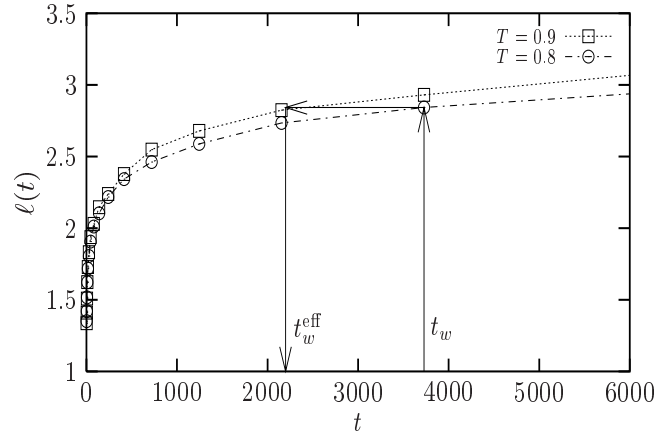,width=9.cm,height=6.5cm} 
\end{center}
\caption{This curve shows that the effective age of the system is 
well described by Eq.~(\ref{estime}).
The parameters are the same as in Fig.~\ref{effective1}.}
\label{check}
\end{figure}

Figure~\ref{check} shows that Eq.~(\ref{estime}) works very well.
Such a relation was proposed in Ref.~\cite{hajime} assuming  
a pure power law growth of the coherence length at all temperatures.
We show below that our data indeed support Eq.~(\ref{estime}), but
is incompatible with a pure power law growth of $\ell_T$.

\begin{table}
\begin{tabular}{c|c|c|c|c} 
$T_2-\Delta T$  &  $T_2$ & $t_w$ & $t_w^{\rm eff}$ &
$t_w^{\rm act}$\\ \hline
0.8 & 0.9 & 1245  & 800  & 615 \\ 
0.8 & 0.9 & 3728  & 2200 & 1629 \\ 
0.8 & 0.9 & 11159 & 7000  &  4317 \\ 
0.7 & 0.8 & 1245  & 650  & 562   \\ 
0.7 & 0.8 & 3728  & 2000 & 1471 \\ 
0.7 & 0.8 & 11159 & 4600 & 3839 \\ 
0.6 & 0.7 & 1245  & 510  & 503 \\ 
0.6 & 0.7 & 3728  & 1450 & 1289 \\ 
0.6 & 0.7 & 11159 & 3700 & 3297 
\end{tabular}
\caption{Effective waiting times for various shift-experiments.}
\label{table2}
\end{table}
 
Different values of $t_w^{\rm eff}$ obtained in a series of shift experiments
are reported in Table~\ref{table2}.
This effective age is also compared to the simple activation prediction,
$t_w^{\rm act}$,
where the same barriers are crossed at the two temperatures. In this case, 
one gets the prediction that
\begin{equation}
\ln \left( \frac{t_w^{\rm act}}{\tau_0} \right)
= \frac{T_1}{T_2} \ln \left( \frac{t_w}{\tau_0} \right).
\label{tweff}
\end{equation}
In this equation, $\tau_0$ is the microscopic time that was extracted 
from the growth of the coherence length at $T=T_c$.
From Table~\ref{table2}, one clearly
concludes that $t_w^{\rm eff} > t_w^{\rm act}$, which suggests
that the microscopic `trial' time is actually much larger than $\tau_0$. 
The same systematic effect has been deduced from recent experiments on 
Ising samples~\cite{Dupuis}.

It is interesting to remark that the simple power law growth $\ell_T \sim
(t/\tau_0)^{1/z(T)}$, with $z(T) = z_c T_c/T$,  
also leads to the purely activated law Eq.~(\ref{tweff}) for the 
effective waiting 
time, and therefore {\it fails} to explain the observed behavior 
of spin glasses.
On the other hand, the mixed critical/activated growth law 
described in Eq.~(\ref{ellt}),
where the microscopic time $\tau_0$ is multiplied by $\ell_T^{z_c}$, 
is indeed  able to account for deviations from Eq.~(\ref{tweff}), as 
already discussed in details in Refs.~\cite{JP,Dupuis}.

We have used the analysis proposed in Ref.~\cite{JP} to extract $\psi$ and
 $\Upsilon_0$
from the data in Table~\ref{table2}. 
Interestingly, we find $\psi \sim 1.1$, $\Upsilon_0 \sim 2.1$, 
compatible 
with the value obtained from the direct fit of the coherence length. 
The agreement between direct and indirect determinations of $\psi$
is an important result of this paper, since it validates the 
analysis performed on experimental data, where no direct determination is 
possible. However, the value of $\psi=1.0$ favoured by our numerical data
is different from the ones reported in previous experimental work on 
Ising spin-glasses using different procedures:
$\psi \sim 0.3 - 0.5$ \cite{Dupuis}, 
$\psi \sim 0.7$ \cite{Suedois},
$\psi \sim 1.9$ \cite{Suedois-Yoshino}.
It is true
that the length scales probed in experiments are at least a
factor ten larger than those probed here.
This does not explain, however, the scattering of the experimental 
data. 

\section{Large temperature shifts: Rejuvenation, Kovacs and memory effects}
\label{largesec}

We turn now to another set of experiments~\cite{cycle3}, 
where larger shifts~\cite{shift3}
$T_1 \to T_2$,
and possibly cycles $T_1 \to T_2 \to T_1$, are performed. 
In the previous section, indeed, the dynamics after a shift 
was the continuation of the aging before the shift.
In this section, we use larger temperature shifts, 
so that the small scale structures that were equilibrated at the first 
temperature have to adapt to the new one. Precisely how this happens is 
what we address
in this section.

\subsection{Is rejuvenation observable in simulations?}

The basic message of large temperature shift experiments is that,
independently of the sign of $T_1-T_2$, aging is `restarted' 
at the new temperature~\cite{shift3}.
This `rejuvenation effect'
can be nicely observed through the measurement of the magnetic
susceptibility $\chi(\omega,t_w)$. For a given frequency $\omega$,
the dominant contribution to the aging part of $\chi(\omega,t_w)$
comes from the modes with a relaxation time $\sim \omega^{-1}$ which are
still out of equilibrium at time $t_w$. Rejuvenation after a negative 
temperature shift comes from fast modes, which were equilibrated at 
$T_1$, but fall out of equilibrium and are slow at $T_2$. Therefore,
one should expect to see this phenomenon if the equilibrium 
conformation of length scales $\lesssim \ell_{T_1} (t_w)$ is 
sufficiently different at the two temperatures (see below for a more
precise statement).

This mechanism is qualitatively different from the 
interpretation involving the notion of temperature chaos~\cite{chaosBM},
and put forward in various approaches~\cite{drop,kohi,hajiji}.
In the latter, the existence of an overlap length $l_o(T_1,T_2)$,
diverging when $T_2 - T_1 \to 0$, is postulated. Its physical content 
is that length scales smaller than  
$l_o$ are essentially unaffected by a 
temperature shift $T_1 \to T_2$, while 
larger length scales are completely re-shuffled by the shift.
In this picture, rejuvenation is thus 
attributed to {\it large} 
length scales. Strong rejuvenation effects therefore 
require a very small $l_o$.

\begin{figure}
\begin{center}
\psfig{file=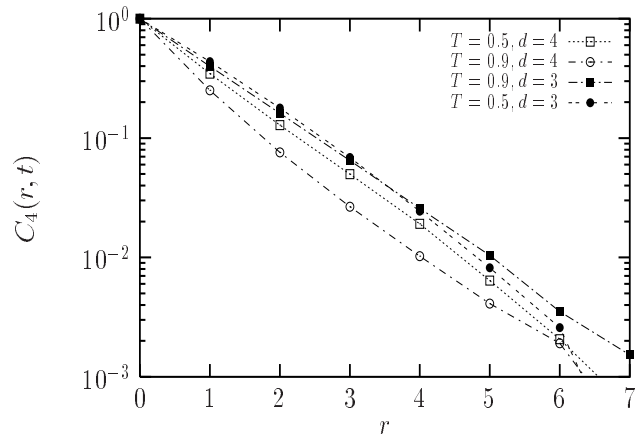,width=9.cm,height=6.5cm} 
\end{center}
\caption{The 4-point correlation
functions at two temperatures in $d=3$ (black symbols) and $d=4$
(open symbols). Times are chosen so that $\ell_{0.9} \sim \ell_{0.5}$, namely
$t(T=0.9) = 720$ and $t(T=0.5)=19307$.}
\label{frozen}
\end{figure}

It turns out that no clear 
rejuvenation effects have ever been observed 
in simulations of the 3 dimensional Ising spin 
glass~\cite{hajime,federico3}.
This was first attributed to the fact that $l_o(T_1,T_2)$ was perhaps 
numerically large, so that no large scale reorganization could be
observed on the time scale of numerical simulations~\cite{hajime}. 
Another possibility is that the Edwards-Anderson model lacks a 
crucial ingredient to reproduce the experiments~\cite{federico3}, or that 
the length and time scales involved in the simulations are too 
small~\cite{federico3}.

From the above discussion, we see that the crucial ingredient 
is the {\it small scale reorganization due to a temperature shift}.
A natural measure of the spatial organization is provided by 
the 4-point correlation, Eq.~(\ref{c4r}).
We show in Fig.~\ref{frozen} the function $C_4(r,t)$
at two different temperatures  $T_1=0.9$ and $T_2=0.5$ in $d=3$ and $d=4$.
Times are chosen so that $\ell_{T_1} \sim \ell_{T_2}$. 
It is clear from this Figure that a temperature shift $T_1 \to T_2$ will 
hardly play any role in $d=3$, whereas the two curves are clearly different 
in $d=4$. 
Another way to see this is to observe the temperature dependence of the 
exponent $\alpha(T)$ reported in Table~\ref{tablealpha}. This exponent is 
almost 
constant in $d=3$, but varies significantly in $d=4$. This observation 
suggests
that no clear effect can be seen in $d=3$, whereas $d=4$ should be more 
favourable. 

\subsection{Negative temperature shifts and rejuvenation}
\label{rejuju}

\begin{figure}
\begin{center}
\psfig{file=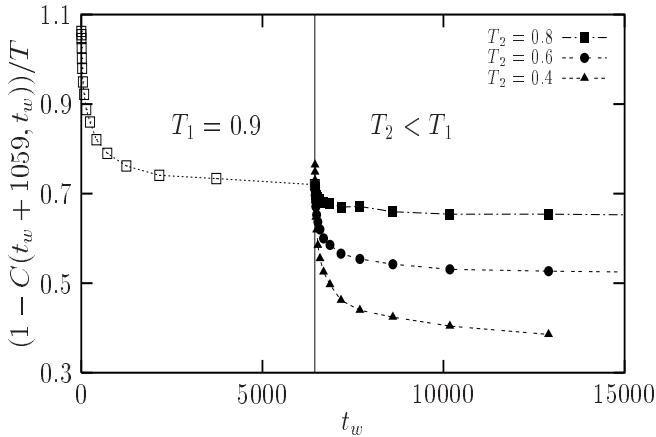,width=9.cm,height=6.5cm} 
\end{center}
\caption{Effect of a negative temperature shift on the `a.c' correlation
function, for $\omega=1/1059$, 
after a long stay $t_s = 6449$ at $T_1=0.9$, for different $T_2$, in $d=4$. 
The larger $\Delta T$, the stronger the rejuvenation.}
\label{chaos}
\end{figure}

This is indeed what we observe numerically
on the analogue of the a.c. susceptibility, defined in Eq.~(\ref{chidef}).
In $d=3$, the amplitude of rejuvenation is very 
small~\cite{hajime,federico3}, as 
expected from the behaviour of the 4-point correlation. In $d=4$, on
the other hand, the a.c. susceptibility  `restarts aging' after a negative
shift $T_1 \to T_2 < T_1$, at time $t_s$, as illustrated in Fig.~\ref{chaos}.
These curves are very similar to what is observed experimentally.

The amplitude of the rejuvenation is found to increase {\it smoothly} with 
the amplitude of the shift $\Delta T = T_1-T_2$. The `temperature chaos' 
picture suggests a
more brutal crossover: no rejuvenation should appear
as long as $l_o > \ell_{T_1}(t_s)$.
There should thus exists a typical shift-amplitude, $\Delta T^*(t_s)$, such 
that for $\Delta T < \Delta T^*(t_s)$,
rejuvenation should be almost absent. 
As discussed in
section~\ref{nochaos}
below, we actually can rule out more directly this interpretation in 
terms of an overlap
length. 

Although a rather strong rejuvenation appears for large $\Delta T$,
one needs to discuss the effect in more details. In particular, the experiments
show that for large $\Delta T$, rejuvenation is `complete' in the sense 
that $\chi(\omega,t_w)$ after the
temperature shift is indistinguishable from the curve obtained after a 
direct quench from high temperatures. This is closely related to the absence 
of cooling rate effects on the a.c. susceptibility, as reported in 
Ref.~\cite{shift2}. 

We have thus compared the evolution
of $\chi(\omega,t_w)$ from our simulation of a temperature shift to the
result obtained after a direct quench. We find that the curves are 
significantly different. The curve after the shift is 
clearly `older' than after a direct quench. 
However, as we discuss in the next section, an experimental quench is
never infinitely fast, contrarily to what can be achieved numerically. 

\subsection{Cooling rate effects and sub-aging}

\begin{figure}
\begin{center}
\psfig{file=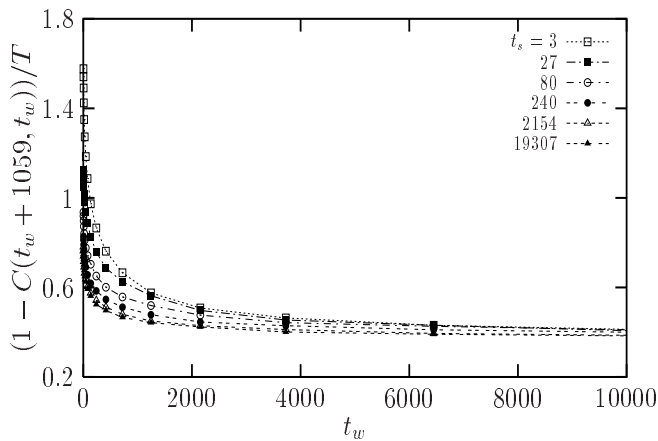,width=9.cm,height=6.5cm} 
\end{center}
\caption{Effect of a negative temperature shift on the `a.c' correlation
function, after a stay at $T_1=0.9$ of various durations $t_s$.
Final temperature is $T_2=0.4$.}
\label{cool2}
\end{figure}

In order to quantify the rejuvenation effect, we
investigate the influence of the time $t_s$ spent at 
$T_1=0.9$ before the temperature shift. 
The evolution of $\chi(\omega,t_w)$ after 
the shift to $T_2=0.5$ for different $t_s$ is shown in Fig.~\ref{cool2}.
We note that as soon as $t_s$ is sufficiently long, $t_s \gtrsim 240$,
the evolution after
the shift becomes independent of $t_s$. 
For smaller $t_s$, on the other
hand, one sees that extra aging contributions are present. 
Hence, for $t_s \gtrsim 240$,
some short scale correlations created at $T_1$ survive at $T_2$,
even for large $\Delta T$, making the relaxation different from what it is 
when $t_s=0$. 
This points towards the absence of temperature chaos and 
will be discussed further in section~\ref{nochaos}.
This shows also that for large enough $t_s$ the system behaves after the 
shift as if 
it had spent an infinite time at $T_1$, i.e. as if $t_s = \infty$. 

The important point now is that experiments always spend some finite time
(actually quite long compared to the microscopic time) at all 
temperatures above the final one $T_2$, where some particularly strong 
correlations very rapidly set in and survive when the temperature is lowered.
Therefore, as
soon as the cooling rate is not extremely fast, the initial configuration
at $T_2$ already has some of the correlations that the system wants to grow
(see also section~\ref{nochaos}). On the other hand, as our simulations show, 
waiting longer at these intermediate
temperatures will not affect further the behaviour at $T_2$. 
The initial age of the system is 
thus effectively non-zero, but very soon independent of the cooling rate.

Interestingly, 
this non-zero initial age induces apparent sub-aging effects. 
Indeed, if $t_{\rm rel}(t_w)=t_0+t_w$, where
$t_0$ approximately 
accounts for the aging accumulated on the cooling path, the effective 
exponent $\mu$ is found to be less than unity:
\begin{equation}
\mu = \frac{d \log t_{\rm rel}}{d \log t_w} = \frac{1}{1+\frac{t_0}{t_w}} < 1.
\end{equation}
We have confirmed this directly on the scaling of the two-time correlation
function $C(t+t_w,t_w)$ obtained after a slow quench in the $d=4$ case. 
We find that $\mu = 0.96 < 1$, whereas the scaling obtained after an
infinitely fast quench indicated super-aging, $\mu = 1.05 > 1$ (see
Fig.~\ref{isocorrresc2} above).
We believe that this effect is significant. 
It is thus tempting to ascribe at least part of the sub-aging effects
seen experimentally to finite cooling rate effects.

\subsection{Temperature cycles and memory}

Since we have considered above both cases of a positive and a negative 
temperature cycle, we are now in position to combine both procedures 
and study temperature-cycles.
The experimental procedure is here $T=\infty \to T_1 \to T_2 < T_1 \to T_1$.
The time spend at $T_1$ is $t_s$ and the time spent at $T_2$ is $t_s'$.
The spectacular `memory effect' arises when the temperature is shifted back 
to $T_1$. It is observed that although aging was fully 
restarted at $T_2$, the system 
has a strong memory of the previous aging at $T_1$. 
The dynamics at $T_1$
proceeds almost as if no cycle to $T_2$ had been performed~\cite{cycle3}.
The coexistence of rejuvenation and memory was made more spectacular 
in the `dip-experiment' proposed in Ref.~\cite{shift2}.
This protocol is too complicated to be studied theoretically, but 
basically carries the same physical content as the cycle we discuss here.

As discussed in \cite{JP,JPB} the memory effect is a simple consequence 
of the separation 
of time and length scales.
When the system is at $T_2 < T_1$, rejuvenation involves very small length
scales as compared to the length scales involved in the aging at $T_1$. 
Thus, when the temperature is shifted back to $T_1$, the correlations 
of length scale $\ell_{T_2}(t_s')$ grown at $T_2$   
almost instantaneously re-equilibrate at $T_1$ (in fact in a `memory' time
scale $t_m$ such that $\ell_{T_1}(t_m) \sim \ell_{T_2}(t_s')$, which 
implies that
$t_m \ll t_s'$ when $\Delta T$ is sufficiently large).
The memory is thus stored in the intermediate length scales,
between $\ell_{T_2}(t_s')$ and $\ell_{T_1}(t_s)$. 

\begin{figure}
\begin{center}
\psfig{file=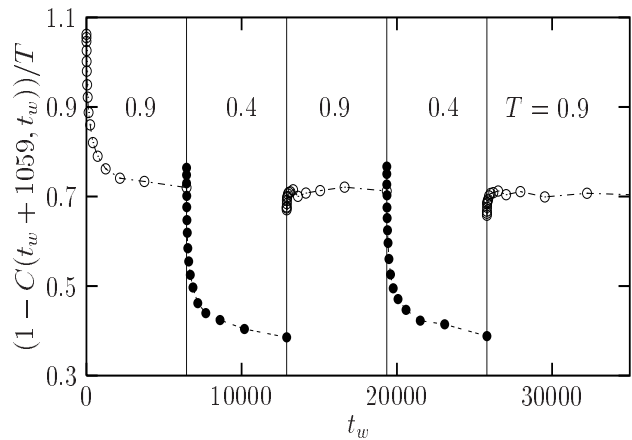,width=9.cm,height=6.5cm} 
\end{center}
\caption{Evolution of the `a.c.' correlation function
in the procedure $T=\infty \to T_1=0.9 \to T_2 =0.4 \to T_1 \to T_2$,
showing, as in experiments the coexistence of rejuvenation
and memory effects.} 
\label{double}
\end{figure}

Hence, the explanation of the memory effect relies on the separation 
of length scales only. This ingredient is distinct from 
the one needed to observe rejuvenation, which relies on the 
reorganization of small length scales after a temperature change.
Since we have shown that 
these two ingredients are present
in the 4 dimensional spin glass, we are able to reproduce
experimental data very well in Fig.~\ref{double}, where, for 
purely esthetic reasons, a double cycle was performed.

\subsection{Positive temperature shifts}

From the results of the previous sections, the physics in a shift experiment 
is the following.
At the first temperature $T_1$, the system evolves towards equilibrium
through the growth of a coherence length $\ell_{T_1}$.
When the temperature is shifted to $T_2$ at time $t_s$, all length scales 
are driven out of equilibrium. Length scales smaller than 
$\ell_{T_1}(t_s)$ undergo a `quench' from $T_1$ to $T_2$, while
larger length scales which were not equilibrated at $T_1$ 
undergo a quench from $T=\infty$ to $T_2$. If $T_2 < T_1$, 
then larger length scales do not matter due to the 
huge separation of time scales. 

The situation
is different in a shift such that $T_2 > T_1$. Then small length 
scales have to `unfreeze' 
to find their new equilibrium at $T_2$, while larger ones which 
where frozen at
$T_2$ grow as if the quench had been from $T=\infty$. 

\begin{figure}
\begin{center}
\psfig{file=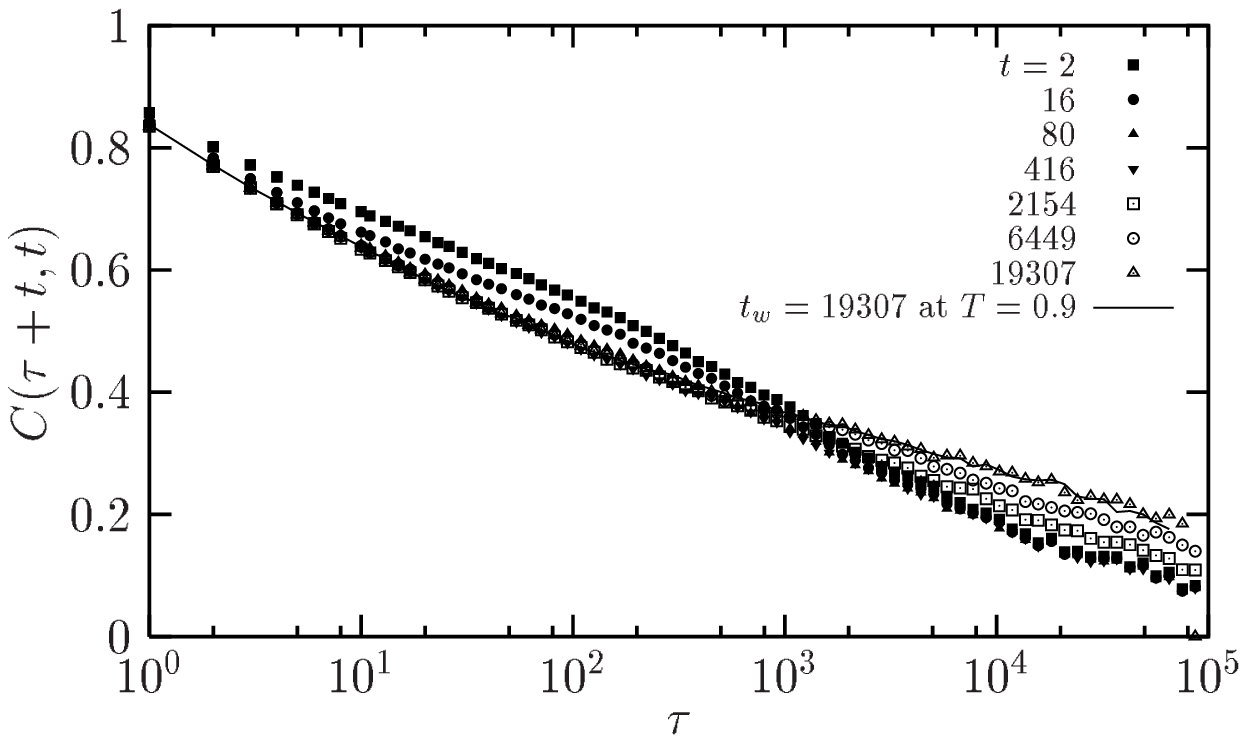,width=9.cm,height=6.5cm} 
\psfig{file=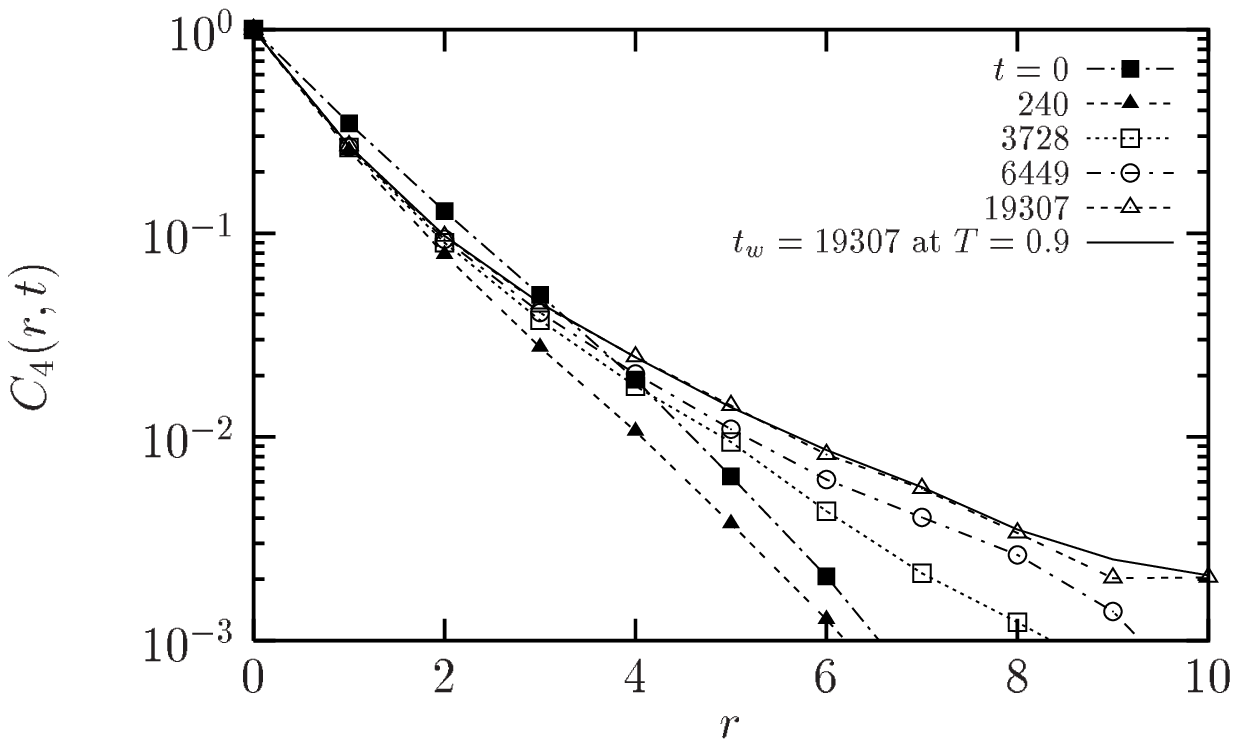,width=9.cm,height=6.5cm} 
\end{center}
\caption{Autocorrelation function (top) and spatial 
correlation function (bottom) in a positive cycle $T_1 = 0.5 \to T_2 = 0.9$.
Times $t$ refer to time spent after the shift.
For comparison, full lines show curves obtained in 
a direct quench to $T_2$. 
Small and large scales behave very differently.
Note also the similarity between both figures
which makes clear the link between time and length.}
\label{poscycle}
\end{figure}

To support further this physical picture, 
we performed a positive temperature shift experiment $T_1=0.5 \to T_2=0.9$, 
after time $t_s=19307$ at $T_1$, and then some extra time $t$ at $T_2$. 
The results are described in Fig.~\ref{poscycle} which shows both the behavior
of the autocorrelation and the 4-point correlation
after the shift. 

Let us comment first on the time correlation functions, Fig.~\ref{poscycle} 
(top).
Immediately after the shift, $t =0$, the decay of the correlation 
shows a short-time part which is slower than the reference curve 
with $t_w=19307$ 
at $T_2$, and a long-time part which is faster. This 
nicely illustrates the two types of structures 
present at that time in the system. 

Then, small scale structures very rapidly equilibrate at the new temperature, 
$t \le 416$. 
Note that it took $t_s = 19307$ to reach the same 
coherence length at $T_1$, a consequence of the length scale separation.
After this short transient, dynamics proceeds as if the initial
stay at $T_1$ was not present, and the subsequent aging is very 
similar to isothermal aging at $T_2$, as soon as $t \ge 2154$.
The same features are also clearly visible on the 4-point correlation 
function, see Fig.~\ref{poscycle} (bottom). Note in particular how small 
scales rapidly `unfreeze' before the large scales evolve towards
equilibrium: the correlation for $t=240$ is {\it below} the one for $t=0$, 
before the coherence length $\ell_{T_2}(t)$ grows beyond $\ell_{T_1}(t_s)$.

\subsection{The `Kovacs effect'}

This dual behavior between small and large length scales 
results in a spectacular effect, which was  
first observed by Kovacs in polymeric glasses~\cite{kovacs}. 
Further developments may be found in 
Refs.~\cite{struik,reviewappl}.
Since it is referred to in the literature as a `memory effect', 
but is different from what the spin glass literature names `memory' 
(see above), we shall follow Ref.~\cite{BerthierEPL} and 
describe this as the `Kovacs effect'.

\begin{figure}
\begin{center}
\psfig{file=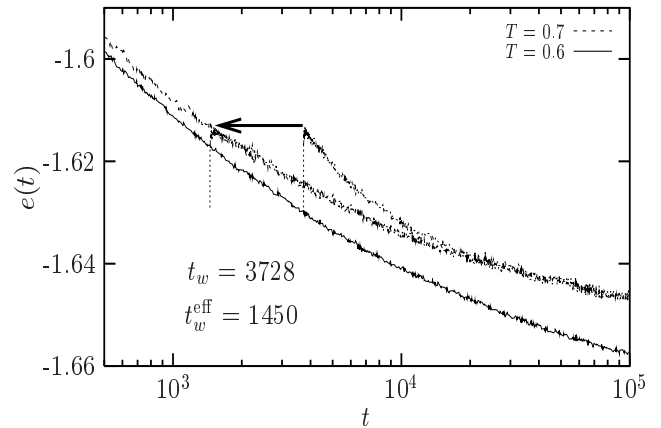,width=9.cm,height=6.5cm} 
\end{center}
\caption{Behavior of the energy density
in a shift experiment from $T_1=0.6$
to $T_2=0.7$, at time $t_w=3728$, compared with its decay in simple aging 
experiments (lines), in $d=3$.
A simple shift with $t_w^{\rm eff}=1450$ allows one to superimpose the curves.}
\label{energy}
\end{figure} 

We focus here on the energy density $e(t)$ after a positive temperature
shift. The Kovacs effect concerns the specific volume, but the difference is 
irrelevant for our purposes. 
Like in recent numerical experiments~\cite{hajime}, 
we find that the decay of the energy density 
following the shift follows the same time evolution as in the simple
aging case, if an appropriate effective waiting time 
$t_w^{\rm eff}$ is properly taken into account, see Fig.~\ref{energy}.
We find that the effective age of the sample defined from the correlation
function or from Eq.~(\ref{estime}) works well for the energy density
also. This is illustrated in Fig.~\ref{energy}.

\begin{figure}
\begin{center}
\psfig{file=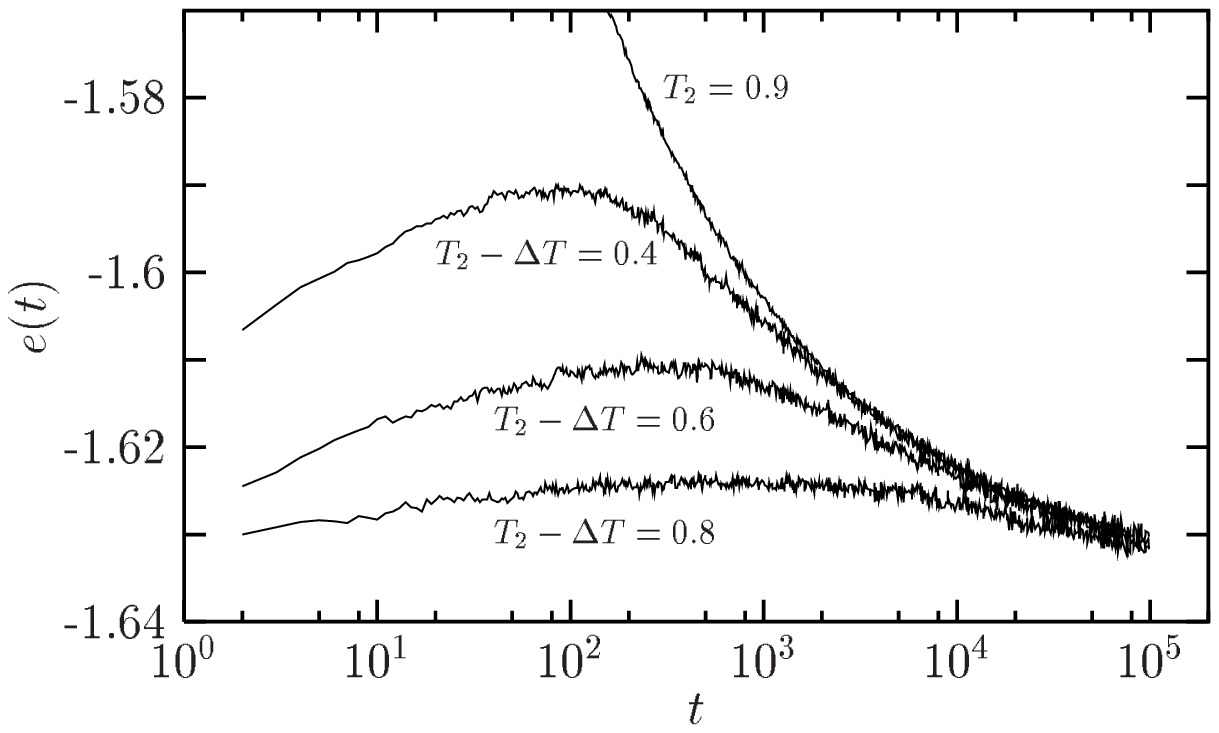,width=8.6cm,height=6.5cm} 
\psfig{file=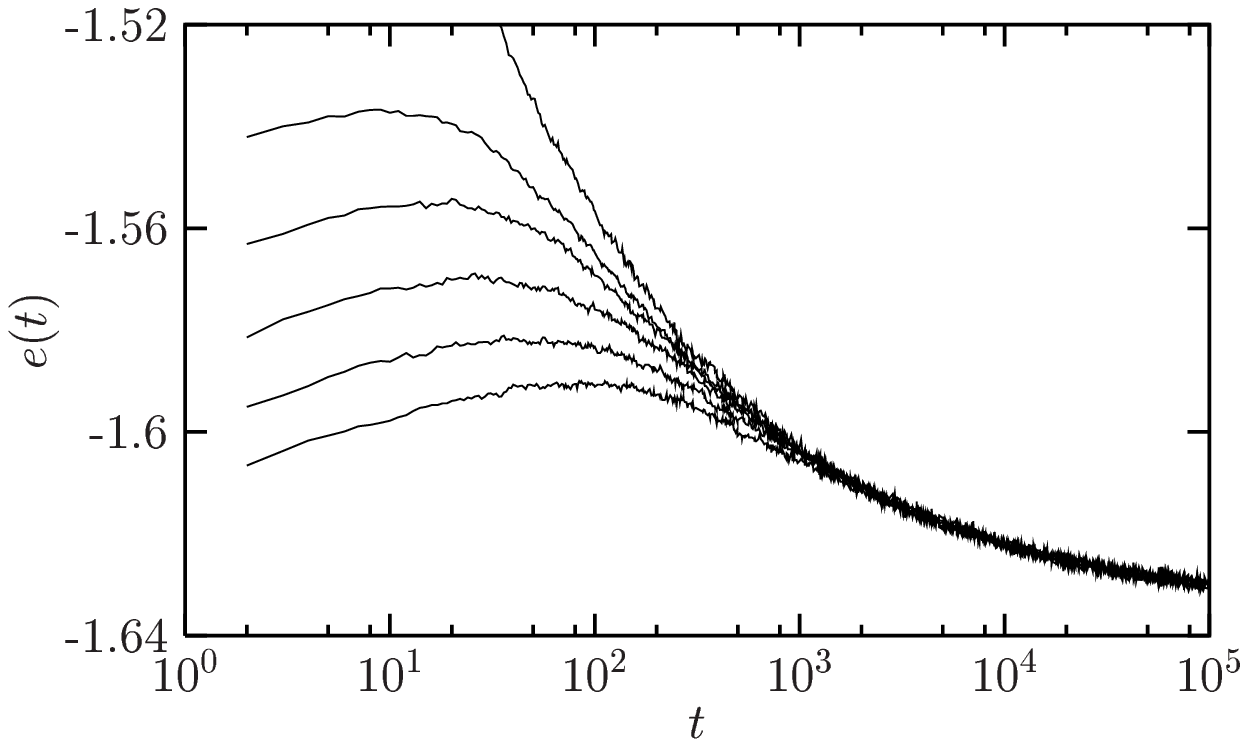,width=8.6cm,height=6.5cm}
\end{center}
\caption{Non-monotonic behavior of the energy in a shift experiment.
Top: Same $T_2$ but different $\Delta T$. Waiting times are such 
that the energy just before the shift is $\sim e_{\rm eq}(T_2)$
as in experiments on polymeric glasses.
Bottom: The $\Delta T$ is the same in all
shifts, but shift times are different, $t_s=57797$, 19307, 6449, 2154, 720 
and $t_s=0$ (from bottom to top).}
\label{kovacs}
\end{figure}

Kovacs~\cite{kovacs} noticed that in a similar protocol on polymer glasses, 
the same non-monotonic initial behavior could be seen in the evolution
of the specific volume as we observe in Fig.~\ref{energy} for the energy 
density of the spin glass immediately after the shift.
Zooming on the
transient region and setting the origin of time when the temperature
is shifted leads to the curves plotted in Fig.~\ref{kovacs}.
The top curves of Fig.~\ref{kovacs} are specially 
designed to follow Kovacs' experiments, where 
the time $t_s$ of the shift is chosen so that 
$e_{T_1}(t_s) = e_{T_2} (t=\infty)$. Since the energy 
density has already the correct equilibrium value at the new temperature, 
the naive expectation is that $e(t>t_s) = const$. 
Instead, the non-monotonic behavior of Figs.~\ref{kovacs}
is observed.

The presence of a growing coherence length allows one to give 
a very simple interpretation of this `Kovacs effect'~\cite{BerthierEPL}.
It results indeed precisely from the dual behavior
of length scales described above. 
When the temperature is shifted to $T_2$, length scales 
shorter than $\ell_{T_1}(t_s)$ have to re-equilibrate at $T_2$, where their
equilibrium energy is higher than at $T_1$. This explains the 
initial rise of $e(t_s+t)$. On the other hand, length scales larger than 
$\ell_{T_1}(t_s)$ 
still have to `cool down' and decrease their energy. These two 
opposite
trends directly explain the `Kovacs effect'. 
This scenario was recently illustrated on the exactly soluble example
the 2D XY model~\cite{BerthierEPL}.

\begin{figure}
\begin{center}
\psfig{file=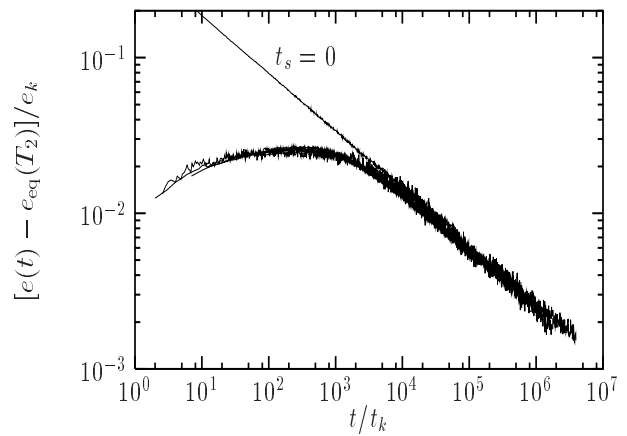,width=8.6cm,height=6.5cm} 
\end{center}
\caption{All the curves
of Fig.~\ref{kovacs} 
are superposed using Eq.~(\ref{scalkoko}).
The curve `$t_s=0$' refers to a direct quench to $T_2$.} 
\label{koko}
\end{figure}

It is possible to be more quantitative here, using the 
coherence length as an ingredient~\cite{BerthierEPL}.
The time scale $t_K$ where the energy density reaches its maximum
corresponds in this picture to the time where small
length scales have re-equilibrated at the new temperature.
Hence, an excellent approximation
for this time scale should be:
\begin{equation}
\ell_{T_2} (t_K) \sim \ell_{T_1}(t_s).  
\end{equation}
This relation says that $t_K$ is an increasing 
function of $t_s$, and a decreasing function of the 
temperature difference, as is obvious from Figs.~\ref{kovacs}. 
The height $e_K$ of Kovacs' hump varies in the opposite direction,
as expected from the inverse power-law dependence of the excess energy 
with the coherence length, $e_K \sim \ell_{T_2}(t_K)^{\theta-d}$, found 
in Ref.~\cite{hajime}. 
We numerically find that all the curves of Figs.~\ref{kovacs}
can actually be collapsed into a single master-curve (see Fig.~\ref{koko}) 
which thus takes the form:
\begin{equation}
e(t+t_s) \simeq e_{\rm eq}(T_2) + e_K  f \left( \frac{t}{t_K}\right),
\label{scalkoko}
\end{equation}
where the scaling function $f(x)$ behaves approximately as a 
power law for large
arguments.

\section{Physical Discussion}
\label{physsec}

\subsection{Physical picture of the spin-glass phase}

Mean-field theories have nothing to say about possible relevant 
length scales (such as $\ell_T$) and their
time and temperature dependence. On the other hand, the droplet picture,
which focuses on relevant length scales, seems to miss some important 
points such as the power-law behaviour of the 4-point correlation function.
This point is important since it
allowed us to account for rejuvenation effects without 
the need of the concept of temperature chaos.

One interpretation of the above results  is that, as 
predicted by mean-field theories and given some credit by recent numerical
work on low-lying excitations~\cite{OMartin,Palassini-Young}, 
different equilibrium 
configurations are accessible to the spin-glass 
in its low temperature phase. These configurations have a global overlap which
is close to zero, but can be locally similar. The fact that the 
stationary part of $C_4(r,t)$ decays as a power-law suggest the existence 
of a fractal `backbone' of spins that have identical mutual orientations for
all these configurations, with a fractal dimension $d_f=d-\alpha$. 
It is reasonable to assume that the small scale properties of this backbone 
will be temperature dependent: more spins will freeze and join the backbone 
as the 
temperature is reduced. The simplest scenario compatible with a zero minimal 
overlap 
is that the backbone is dense on small scales, and fractal on large scales, 
with a 
temperature dependent crossover length $\ell^*(T)$. The effective exponent 
$\alpha(T)$
would in this case decrease with temperature, as seen numerically. Another 
possibility
is that the fractal dimension (and thus the exponent $\alpha$) is truly 
temperature
dependent, as in the low temperature phase of the 2D XY model. As 
discussed above,
there are actually many phenomenological similarities between 
the spin-glass 
phase and the
2D XY model,
provided length scales, rather than time scales, 
are compared~\cite{BerthierEPL,behose}. 

The power-law decay of $C_4(r,t)$ suggests that the whole spin-glass phase is 
in a certain sense critical, at least in the `zero-overlap' sector which 
was indeed
found to be mass-less in replica field 
analysis~\cite{Cirano}. However, this 
is 
{\it not} in contradiction 
with the existence of a finite correlation length 
$\xi(T)$ 
separating critical from activated dynamics within a single `state'. A pure
power-law growth of $\ell_T$ is not necessarily a consequence of the 
criticality 
of the spin-glass phase. 

Of course, the numerical evidence for this scenario is fragile, and it
could be that $C_4(r,t \to \infty)$ in fact tends for large $r$ towards 
$q_{\rm EA}^2$. For the purpose of interpreting aging experiments, however, 
it is sufficient that this scenario holds even approximately on the relevant 
time and length scales. 

\subsection{Rejuvenation from small scales and absence of temperature chaos}
\label{nochaos}

The role of temperature changes can be exactly computed in the Random Energy 
Model~\cite{MartaEPL}
and in the critical phase of the XY model~\cite{BerthierEPL}. 
Both show that 
it is possible to
induce strong rejuvenation effect without the existence of an overlap length. 
Several
facts, reviewed in Ref.~\cite{JPB}, also suggest that the 
overlap length is not relevant to the experimental findings. 
Here, we want to address this question 
more precisely 
on the basis of numerical results. As shown above, we can now observe beyond 
any doubts 
rejuvenation (and memory) effects in the $d=4$ Edwards-Anderson model which 
are very 
similar to those observed experimentally. We have also investigated directly 
the way 
configurations evolve during a temperature shift using a mixed 4-point 
correlation 
function, defined as follows:
\begin{equation}
C_4 (r,\ell,T_1,\Delta T) = \frac{1}{N}
\sum_{i=1}^N  \overline{ \langle s_i^a(t_w) s_{i+r}^a(t_w) 
s_i^b (t_w') s_{i+r}^b(t_w') \rangle},
\end{equation}
where replica $a$ is at temperature $T_1$, replica $b$ at temperature 
$T_2=T_1-\Delta T$,
and the times $t_w$, $t_w'$ are chosen such that the coherence length is 
equal to a common value 
$\ell$ at the two temperatures. Obviously, when $\Delta T=0$, this 
correlation function is
identical to the previous one. For $\Delta T > 0$, this correlation function 
measures the
similarity between the patterns grown at the two different temperatures. In a 
temperature chaos 
scenario, one expects the following inequality:
\begin{equation}\label{bound}
C_4 (r,\ell,T_1,\Delta T) \leq C_4 (r,\ell,T_1,\Delta T=0).
\end{equation}

\begin{figure}
\begin{center}
\psfig{file=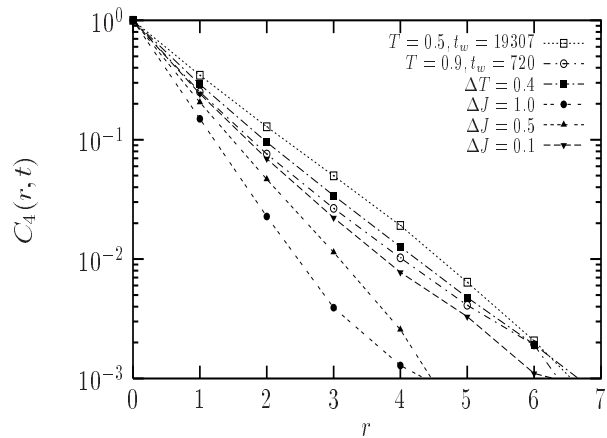,width=8.6cm,height=6.5cm} 
\end{center}
\caption{Mixed correlation functions, both for temperature changes and for 
coupling changes (black symbols), compared to standard four-point
functions (open symbols). It is clear that 
the two perturbations ($\Delta T$, $\Delta J$) 
have qualitatively different effects.}
\label{mixt}
\end{figure}

Figure~\ref{mixt} shows that this is not the case. The results are actually 
compatible
with the idea that the same patterns grow at the two temperatures -- the 
backbone 
supporting the common parts of these patterns being more fluffy at lower 
temperatures.
This conclusion was already reached above when we discussed cooling rate 
effects.

An interesting comparison can be made with a situation where chaos {\it is} 
expected, e.g.
when the couplings are changed~\cite{chaosBM}. 
We therefore also show in the same figure the 
mixed 
correlation function when the couplings are changed between replica $a$ and 
replica 
$b$ according to:
\begin{equation}
J_{ij} \longrightarrow \frac{J_{ij}+\Delta J_{ij}}{\sqrt{1+\Delta J^2}},
\end{equation}
where $\Delta J_{ij}$ are independent Gaussian variable of variance $\Delta 
J^2$ and mean 0. 
In this case, the inequality (\ref{bound}) is indeed clearly observed. 
We conclude thus that $\Delta T$ and $\Delta J$ have 
a qualitatively different influence on the system.

Note that our mixed correlation function, once integrated over space, leads 
to the
overlap between the two temperatures. The latter quantity was studied directly 
in Ref.~\cite{Billoire} in $d=3$, with conclusions similar to ours.

The {\it simultaneous} observation of rejuvenation and absence of temperature 
chaos is an important result of this paper. 
In $d=3$, no temperature chaos was found, but no rejuvenation 
either. This
left the door open to the possibility that the length scales investigated 
were too small
to observe these two effects.
We have thus demonstrated that both issues can be separated. Of course 
temperature chaos on large length scales is still possible, but is not
needed to interpret rejuvenation effects.

In summary, our results confirm that rejuvenation is due to the freezing of 
small length scale 
modes which were `fast' at the higher temperature. This freezing changes the 
correlations
on small scales, as seen on the 4-point correlation function. This is in 
agreement with the
scenario based on a hierarchy of length scales proposed in 
Ref.~\cite{JP,JPB}, and 
with the
phenomenology of the XY model~\cite{BerthierEPL}, and is markedly different 
from the temperature chaos picture.
This feature can be illustrated 
in the 2D XY model~\cite{peter}, 
where each Fourier mode $\varphi(q)$ of the order parameter
is affected by a temperature shift. In the spin-wave approximation, 
one has $\langle \varphi(q) \rangle \sim T/q^2$. 
Hence, {\it each} mode is affected when 
the temperature is changed by $\Delta T$ by an amount 
\begin{equation}
\delta \langle
 \varphi (q) \rangle  \sim \frac{\Delta T}{q^2},
\end{equation}
which shows that larger length scales are more influenced, but with 
no typical `overlap length'.

\section{Summary and conclusion}
\label{discussion}

The interest of a long paper is that a detailed discussion of rather subtle 
points 
can be given. The drawback, obviously, is that the message is somewhat 
diluted. We 
therefore give in this last section the main conclusions from our study and 
end on 
open problems.

\begin{itemize}
\item Aging dynamics in spin-glasses can be associated with the growth of a 
coherence 
length $\ell_T$, separating small, equilibrated scales $< \ell_T$ from large 
frozen,
out of equilibrium scales $> \ell_T$. This scale is however not a domain size 
in the
usual sense, but rather the size of a backbone of spins common to all 
spin-glass configurations.
This interpretation stems from the power-law decay of the 4-point correlation 
function 
from which $\ell_T$ is extracted.

\item The coherence length $\ell_T$ follows a critical power-law growth at 
small times 
that becomes activated for larger times, and is well described by Eq. 
(\ref{ellt}). 
The associated barriers $\Upsilon(T)$
vanish at the critical temperature. The barrier 
exponent $\psi$ 
was estimated to be $\psi \sim 1.0$ for $d=3$ and 
$\psi \sim 2.3$ in $d=4$. 

\item This mixed critical/activated growth law allows one to interpret 
several important
aspects of both simulations and experiments, for example the deviations from 
a purely 
activated behaviour that are revealed by temperature shift procedures, or the 
super-aging
behaviour of the correlation function observed in $d=4$.

\item The short scale behaviour of 
the 4-point correlation is quite sensitive to temperature in $d=4$, but much 
less
in $d=3$. This in turn leads to strong rejuvenation effects in $d=4$, quite 
similar to
those observed in experiments, that we observe for the first time in 
simulations.

\item An interpretation of the observed rejuvenation in terms of temperature 
chaos is, 
we believe, ruled out: see Fig. (\ref{mixt}). Rather, some correlations built 
at a higher temperature persist and are
reinforced at lower temperatures. 

\item A finite cooling rate effect follows from 
this, which,
interestingly,
leads to an apparent {\it sub-aging} behaviour for the correlation function, 
instead
of the super-aging that holds for an infinitely fast quench. The
cooling rate dependence however saturates quickly as soon as the cooling rate 
is not 
infinitely fast. Both these features agree with experiments, for which the 
cooling rate
is always very slow compared to microscopic frequencies.

\item The dichotomy between small, equilibrated scales and large, frozen 
scales allows 
one to account semi-quantitatively for many features, such as the role 
of temperature shifts, the 
memory effect or the Kovacs' hump. 

\end{itemize}

Although our results are suggestive, several unsettled points remain. In 
particular, 
rejuvenation effects are found in $d=4$, but not in $d=3$, whereas 
experiments are 
obviously performed in $d=3$. We conjecture that for the time scales 
investigated, the
large scale topology of space is irrelevant, and the major difference between 
$d=3$ and
$d=4$ should rather come from the local connectivity. Hence it should be 
possible to obtain 
rejuvenation in a $d=3$ model with more neighbours, and reproduce most 
experimental results with a realistic model.

The most important 
theoretical point is obviously the nature of the spin-glass phase. A
well posed problem (but very difficult to settle numerically) is 
the true long distance behaviour of the 4-point correlation function:
power-law decay, as expected from replica symmetry breaking theories, or 
convergence 
towards $q_{EA}^2$, as for a disguised ferromagnet? The final picture of real 
spin-glasses
might in the end have borrow concepts from both theories. The hope is that
the concepts that will emerge will be useful to understand many other
glassy systems, which share a very similar phenomenology.

\section*{acknowledgments}
We thank L.~Bocquet, V.~Dupuis, J.~Hammann, P.~Holdsworth,
O.~Martin, M.~M\'ezard, M.~Ocio,
F.~Ricci-Tersenghi, F.~Ritort, M.~Sales, E.~Vincent, H.~Yoshino, 
and P. Young for useful discussions. 
This work is supported  
by the P\^ole Scientifique de Mod\'elisation Num\'erique
at \'Ecole Normale Sup\'erieure de Lyon.
L. B. would like to thank Marin Berthier and Constant Berthier
for their (noisy) support during the preparation of the manuscript. 

\section*{Note added}

After this manuscript appeared as a preprint (condmat/0202069), 
a paper by Yoshino et al. (cond-mat/0203267) appeared where the 
dynamics of 4-d EA model is studied, and the results
also interpreted as a crossover between critical and activated 
dynamics. The value of the
exponents $z_c$ and $\psi$ given in that paper slightly differ 
from those obtained here.
For example, $\psi$ is found to be in the range $2.5-3$ whereas 
we report $\psi \sim 2.3$.
One possible explanation is that the procedure to extract 
$\ell_T(t_w)$ form $C_4(r,t_w)$ 
is quite different.

\end{document}